\documentstyle[amssym]{mn}
\input{epsf.sty}
\input{psfig}
\def\pageoffset#1#2{\hoffset=#1\relax\voffset=#2\relax} 
\pageoffset{0pc}{2.5pc} 
 at 10truept

\def\and  {\it {et al.} \rm}

\def\etal{{\rm et~al. }}
\def\spose#1{\hbox to 0pt{#1\hss}}
\def\simlt{\mathrel{\spose{\lower 3pt\hbox{$\mathchar"218$}}
     \raise 2.0pt\hbox{$\mathchar"13C$}}}
\def\simgt{\mathrel{\spose{\lower 3pt\hbox{$\mathchar"218$}}
     \raise 2.0pt\hbox{$\mathchar"13E$}}}
\def\be{\begin{equation}}
\def\ee{\end{equation}}
\def\bce{\begin{center}}
\def\ece{\end{center}}
\def\bea{\begin{eqnarray}}
\def\eea{\end{eqnarray}}
\def\ben{\begin{enumerate}}
\def\een{\end{enumerate}}

\def\ni{\noindent}

\def\brr{\begin{array}}
\def\err{\end{array}}

\def\etal{{\rm et~al. }}

\def\nh1{n_{\rm HI}}

\def \p1dk {P_{\rm 1D}(k)}
\def \simlt {\mathrel{\spose{\lower 3pt\hbox{$\mathchar"218$}}
     \raise 2.0pt\hbox{$\mathchar"13C$}}}
\def \simgt {\mathrel{\spose{\lower 3pt\hbox{$\mathchar"218$}}
     \raise 2.0pt\hbox{$\mathchar"13E$}}}

\def \cS {{\cal S}}
\def \P {{\cal P}}
\def \L {{\cal L}}
\def \a {{\bf a}}
\def \M {{\bf M}}

\def \vr {{\bf r}}
\def \vv {{\bf v}}
\def \vH {{\cal H}}
\def \vP {{\bf P}}

\def \be {\begin{equation}}
\def \en {\end{equation}}
\def \bea {\begin{eqnarray}}
\def \ena {\end{eqnarray}}

\def \bi {\begin{itemize}}
\def \ei {\end{itemize}}

\def \etal {{\it et al. }}

\begin{document}
\title[Skeleton as a probe of the cosmic web]{
Skeleton as a probe of the cosmic web: the 2D case}

\author[D. Novikov, S. Colombi \& O. Dor\'e ]
{\ni Dmitri Novikov,$^1$ St\'ephane Colombi,$^2$ Olivier Dor\'e$^{2,3}$ \\
$^1$ Astrophysics, University of Oxford, Denys Wilkinson Building, Keble Road, Oxford OX1 3RH, England\\
$^2$ Institut d'Astrophysique de Paris, 98 bis boulevard Arago, 75014 Paris, France \\
$^3$ Department of Astrophysical Sciences, Princeton university, Peyton Hall, Ivy Lane, Princeton, NJ 08544, USA\\ 
novikov@astro.ox.ac.uk, colombi@iap.fr, olivier@astro.princeton.edu
}
\maketitle

\begin{abstract}
We discuss the skeleton as a probe of the filamentary structures of a 2D random field. 
It can be defined for a smooth field as the ensemble of pairs of field lines 
departing from saddle points, initially aligned with
the major axis of local curvature and connecting them to local maxima. 
This definition is thus non local and makes analytical predictions difficult, so
we propose a local approximation: 
the {\em local} skeleton is given by the set of points 
where the gradient is aligned with the local curvature major axis and 
where the second component of the local curvature is negative. 

We perform a statistical analysis of the length of the {\em total} local skeleton, chosen
for simplicity as the set of all points of space where the gradient is either parallel or
orthogonal to the main curvature axis. In all our numerical experiments, which
include Gaussian and various non Gaussian realizations such as $\chi^2$ fields and
Zel'dovich maps, the differential length is found within a normalization factor
to be very close to the probability distribution
function of the smoothed field. This is in fact explicitly demonstrated in the Gaussian case.

This result might be discouraging for using the skeleton as a probe of non Gausiannity, but our 
analyses assume that the total length of the skeleton is
a free, adjustable parameter. This total length could in fact be used to constrain cosmological
models, in CMB maps but also in 3D galaxy catalogs, where it estimates
the total length of filaments in the Universe. Making the link with other works,
we also show how the skeleton can be used to study the dynamics of large scale structure.
\end{abstract}


\section{Introduction}

The observed large scale distribution of galaxies presents remarkable structures, 
such as clusters of galaxies, filaments, sheets 
and large voids. It is widely admitted that these structures grew from small initial
fluctuations through gravitational instability. At very large scale, the
filamentary pattern seen in the cosmic web is expected to be similar to that of the initial
field (e.g., Bond, Kofman \& Pogosyan 1996). 
Since these primordial inhomogeneities also imprinted the temperature fluctuations
seen now in the Cosmic Microwave Background (CMB), the characterization of the
observed large scale structures both in galaxy catalogs and in CMB maps can help
to probe the nature of these primordial fluctuations, in particular whether
they are Gaussianly distributed or not. Furthermore, a rigorous topological description
of the observed structures is necessary to constrain efficiently models of large scale structure.
For instance, a precise definition is needed for clusters of galaxies before inferring any constraints 
from their studies, e.g density and temperature profiles but also luminosity function and clustering. 
Analogously, a precise and practical definition of filaments would allow us to use them similarly as cluster
of galaxies.

Various methods have been proposed to characterize the morphology of large
scale structures. In general, one studies the topological properties of excursion sets,
i.e. regions below or above a density threshold. The statistics most explored
up to know are the genus or the closely related Euler characteristic 
(see, e.g., Doroshkevich 1970; Gott, Melott \& Dickinson 1986),
the more complete set of observables given by Minkowski functionals
(see, e.g., Mecke, Buchert \& Wagner 1994)
 and related statistics such as shape finders (e.g., Sahni, Sathyaprakash \& Shandarin 1998),
but also estimators based on percolation analysis
(see, e.g., Zel'dovich 1981; Zel'dovich, Einasto \& Shandarin 1982; Shandarin 1983;
Bhavsar \& Barrow 1983)
and minimum spanning tree construction (e.g., Barrow, Bhavsar \& Sonoda
1985),
such as for example 
structure functions or related shape estimators based on moments of inertia 
(see, e.g., Babul \& Starkman 1992).

In general, topological descriptors are primarily used to constrain the level 
of non Gaussianity in the sample, since there often exists analytical predictions 
in the Gaussian case (see, e.g., Doroshkevich 1970 for the genus; 
Tomita 1986 for the Minkowski functionals). 
To do so, other estimators exist also, based on spatial correlation analysis, 
such as higher-order correlation functions
in real or Fourier/harmonic space (e.g., Peebles 1980 and references therein),
the probability distribution function (pdf) of the smoothed density field and its moments
(e.g., Bernardeau et al. 2002 and references therein), 
higher order moments the wavelet coefficients (e.g., Aghanim \& Forni 1999; 
Hobson, Jones \& Lasenby 1999), 
peak and excursion set subcomponents statistics (see, e.g., Bardeen et
al. 1986; Bond \& Efstathiou 1987 and Dor\'e \etal 2003 for a somewhat
related statistic),
phase correlation analysis (e.g., Chian \& Coles 2000; Naselsky, Novikov \& Silk 2002), etc. 
In principle, all these statistics combine the data
in very specific ways, so they altogether provide complementary analysis of the data.\footnote{see, 
e.g., Shandarin 2002 for a comparison of the pdf and the Minkowski functionals
as estimators of non Gaussianity in 2D maps; Phillips \& Kogut 2001 for a comparison
between genus,  extrema correlation function and bispectrum; Barreiro, Mart\'{\i}nez-Gonz\'alez \& Sanz 2001
for a comparison between number density, eccentricity and Gaussian curvature of hot spots, and genus as
estimators to probe non-Gaussianity in CMB samples.} 
However, at variance with traditional statistical estimators, 
topological estimators help as well to characterize in a very intuitive way
the topology of structures in terms of filaments, sheets, clusters  and 
voids quantitatively. For instance, the
Minkowski functionals provide a complete basis of simple estimators to estimate
the morphology of an excursion set (see, e.g., Kerscher 2000 for a review on the subject), 
e.g. its degree of compacity and of filamentarity. 

In this paper we focus on the skeleton, which aims at extracting from the cosmic
web its filamentary pattern. More specifically, the goal is to draw in the observed
structure a set of lines which reproduces well the filamentary pattern guessed by eyes.
A natural tool to do so in set of points such as galaxy catalogs is the minimum spanning tree
(e.g., Barrow, Bhavsar \& Sonoda 1985). 
It is a connected structure superposed to the set of points, with no loop and which is
the shortest possible. Of course, using the minimum spanning tree as such is not very
helpful since it is highly irregular and it is difficult to establish a link to
the large scale features of interest, but there are technics to filter small scale
noise consisting in ``pealing'' the tree, i.e. removing from it short branches. 
Even if it is successful in extracting the main filamentary features from the catalog,\footnote{see,
e.g. Fig.~1 of Doroskevich et al. 2001 for a nice illustration on the Las Campanas Redshift Survey.}
the minimum spanning tree remains by nature unsmooth and difficult if not impossible
to manipulate in order to perform analytic calculations. 

The technique we aim to
employ in this paper to extract the skeleton from the data sample 
is completely different and relies on Morse theory (see, e.g., Milnor 1963; 
Colombi, Pogosyan \& Souradeep 2000; Jost 2002). 
It requires the field to
be sufficiently differentiable and non degenerate as explained more in details 
later and thus some smoothing with e.g. a Gaussian window of the data
file is necessary to use such technique.\footnote{note that the concept of smoothing introduce a scale
in the problem: smoothing at different scales will not produce the same skeleton but
will have interesting links to dynamics as discussed in the end of this paper.}
\footnote{Hence, at variance with the minimum 
spanning tree method, which has the advantage
to deal directly with the discrete nature of galaxy sample, our method will be difficult
to apply to real galaxy catalogs unless smoothing at sufficiently large scales.}

We shall see that the skeleton can then be
then rigorously defined as a set of pairs of special field lines departing from saddle points. 
The problem is that the skeleton defined as such is non local: indeed, to draw any
field line, one has to resolve the trajectory of a particle following the equation
of motion given by $d\vr/dt=\nabla \rho$. This non local nature of the skeleton makes
analytic predictions rather difficult. Furthermore, as discussed in Appendix A, 
it is difficult to find a reliable algorithm to draw it. 

The main points of this paper, which focusses on the 2D case, are the following: (i)
find a local approximation of the skeleton to address the issues just raised above, 
(ii) test this local approximation as a statistical
tool to probe non Gaussian features of the density field in e.g. CMB maps, (iii) establish the link to dynamics.
The last point will be only treated superficially through simple illustrative examples, relying
mostly on the Zel'dovich approximation (Zel'dovich 1970), since it is clearly more interesting to treat it in detail
in the 3D case.\footnote{Note however, that dynamics in 2D can be still of interest, particularly
in relation to reconstruction of the projected mass distribution in weak lensing experiments: for instance
the skeleton can be used as a tool to test the quality of the reconstruction methods.} 

This paper is thus organized as follows. In
Section~\ref{sec:skedef2d}, We define the skeleton 
in the framework of Morse theory and discuss some of its properties. 
We find a local approximation to it, relying on two independent methods.
This {\em local} skeleton is contained in the set
of points of space where the gradient of the density field is aligned with one
axis of the local curvature, that we call the {\em total} local skeleton.
We show through examples that the local skeleton and the real
skeleton are quite alike, both by visual
inspection and by comparing their respective lengths and we discuss the differences found. 
In Section~\ref{sec:stat}, we study the differential length of the total local skeleton as a function
of density threshold. We find experimentally that it scales very much like the pdf 
of the smoothed density field, a property that we demonstrate explicitly in the Gaussian case.
Finally, Section~\ref{sec:discu} discusses the results and makes the link to dynamics. 
An extensive appendix discusses the numerical calculations of this paper, which were
performed with a dedicated package.

\section{The skeleton of a 2D random field}
\label{sec:skedef2d}
In a two-dimensional field, one would naturally define the skeleton  as
a set of ridges connecting local maxima and separating under-dense
regions. In what follows, we first give the practical mathematical
definition corresponding to this view (\S~\ref{sec:definition}).
It is shown that the skeleton is a set of special field lines,
i.e. a particular set of curves parallel to the gradient of the field,
and passing through maxima and saddle points. However this definition
is not very practical at least from the theoretical point of view,
because it is non local and makes analytic predictions
difficult.  To enforce locality, we define an other
set of curves that is aimed to be close to the real skeleton
(\S~\ref{sec:localapp}). 
\begin{figure*}
\centerline{\hbox{
\psfig{file=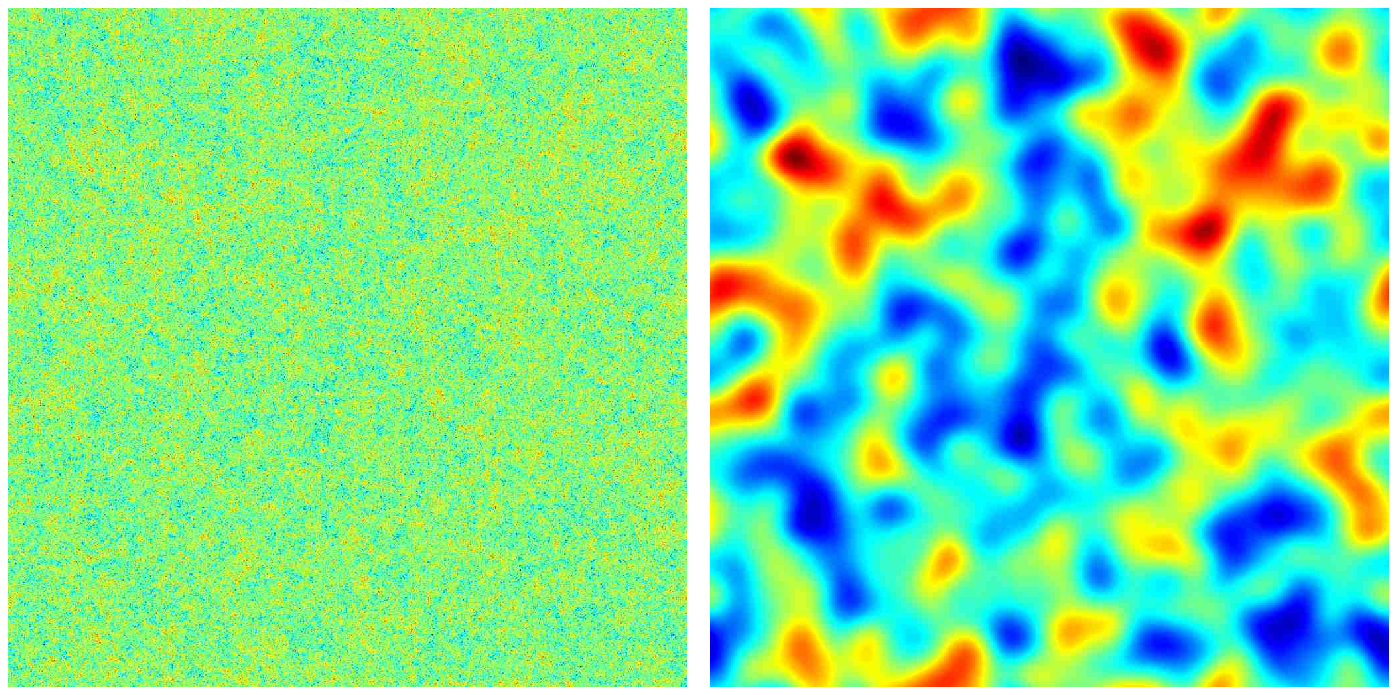,bbllx=116pt,bblly=562pt,bburx=523pt,bbury=763pt,width=12.5cm}
}}
\caption[]{An example of Gaussian field and its smoothed counterpart.

{\em Left panel:} a periodic realization of a 2D Gaussian random
field on a grid of size $1024\times 1024$ pixels with
a scale-free power-spectrum, verifying $P(k) \propto k^{n}$ with
$n=-1$. 

{\em Right panel:} the same field smoothed with a Gaussian window of
radius 25 pixels.}
\label{figure0}
\end{figure*}
To do so we use two different approaches, which actually end in the same
definition for the local skeleton. The first one consists 
in Taylor expanding the special field lines in the neighborhood of 
saddle points and local maxima while the second one consists 
in finding  points along isocontour lines such that the gradient of the
density field has extremal magnitude. 
In \S~\ref{sec:illustration}, our arguments will be illustrated
by practical examples on a Gaussian field and its Zel'dovich mapping.
 As a probe of the local skeleton, in
addition to visual inspection, we shall compare, for the examples
considered here, its length as a function of density threshold with the
length of the real skeleton.

In what follows, we consider a 2D random field, $\rho(\vr)$, e.g. a temperature
map of the CMB. To assume sufficient differentiability we convolve it with a gaussian window
of size $\ell$, as illustrated by Fig.~\ref{figure0}.
The smoothed field, still noted $\rho$, is furthermore 
supposed to be sufficiently non degenerate, and in particular has the
following properties:
\begin{enumerate} 
\item Its gradient cancels in a discrete set of
critical points, which can be separated into three subclasses, local
minima, local maxima and saddle points;
\item The eigenvalues $\lambda_1 \geq
\lambda_2$ of its Hessian, $\vH \equiv \partial^2\rho/\partial r_i \partial
r_j$, verify the following properties: the regions of space
where $\lambda_1=0$
or where $\lambda_2=0$ are sets of smooth curves never passing through
critical points. The intersection of these two sets of curves, where
$\lambda_1=\lambda_2=0$, is therefore a discrete set of points not
containing any critical 
point.
\end{enumerate}
Given these last definitions for $\lambda_1$ and $\lambda_2$, 
the local maxima, saddle point and local
minima verify respectively  
$0 > \lambda_1 \geq \lambda_2$, $\lambda_1 > 0 > \lambda_2$
and $\lambda_1 \geq \lambda_2 > 0$. 
\subsection{Definition}
\label{sec:definition}
We first define the peak patches (void patches) as the regions of
space containing all the points
converging to the same local maximum (local minimum)
while going along the field lines in the direction (opposite direction)
of the gradient,  $\nabla \rho\equiv \partial \rho/\partial r_i$. 

\begin{figure*}
\centerline{\hbox{
\psfig{file=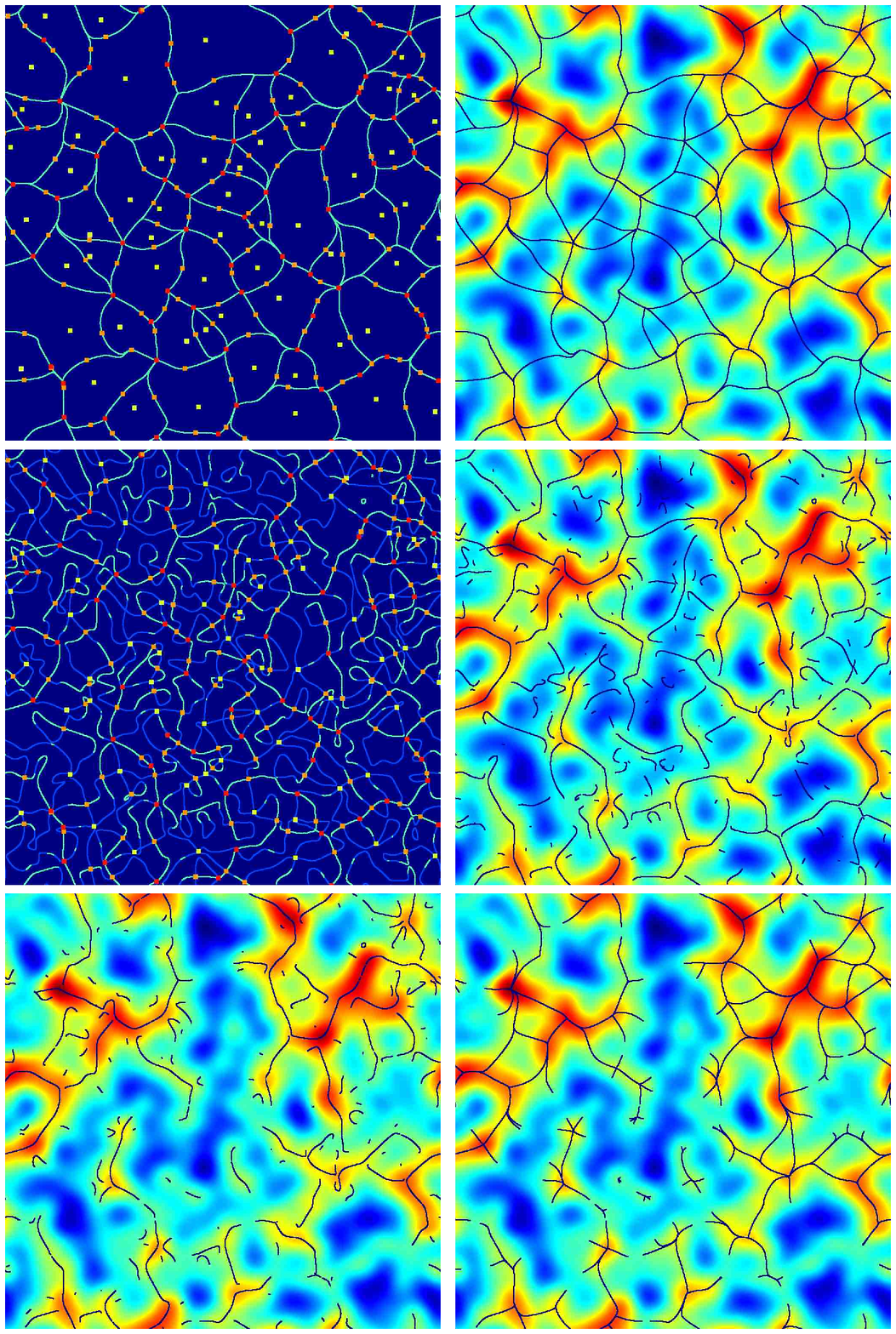,bbllx=116pt,bblly=159pt,bburx=523pt,bbury=763pt,width=12.5cm}
}}
\caption[]{Skeleton and its local approximation for the Gaussian field
of Fig.~\ref{figure0}. 

{\em Upper left panel:} the skeleton is drawn
as well as the critical points: local minima in yellow, saddle points
in orange and local maxima in red. As discussed in the text, the
skeleton passes through all the maxima and the saddle points. The
local maxima are the nodes where several lines converge, while
the saddles points have only one line passing through. Note as well
that local maxima are always connected to saddles and reciprocally,
except in e.g. the lower left of the panel, when one can see three
saddle connected to each other. This configuration is theoretically
forbidden unless there is some degeneracy in the field, that we
suspect is due to our numerical implementation, as further discussed
in Appendix. 

{\em Upper right panel:} the skeleton is superposed to
the smoothed field. 

{\em Middle left panel:} same as for the upper
left panel, but for the local approximation of the skeleton. The dark
plus light blue lines assume ${\cal S}=0$ [eq.~(\ref{eq:localsketot})],
while the light blue lines verify the more constraining conditions
given by eqs.~(\ref{eq:lambpos}) and (\ref{eq:eigen}). 

{\em Middle right panel:} same as upper right panel but for the local
approximation of the skeleton. 

{\em Lower left and lower right panels:} the local approximation and
the real skeleton are again superposed to the smooth field, but
restricted to over-dense regions $\rho \geq \langle \rho \rangle$.}
\label{figure1}
\end{figure*}
\begin{figure*}
\centerline{\hbox{
\psfig{file=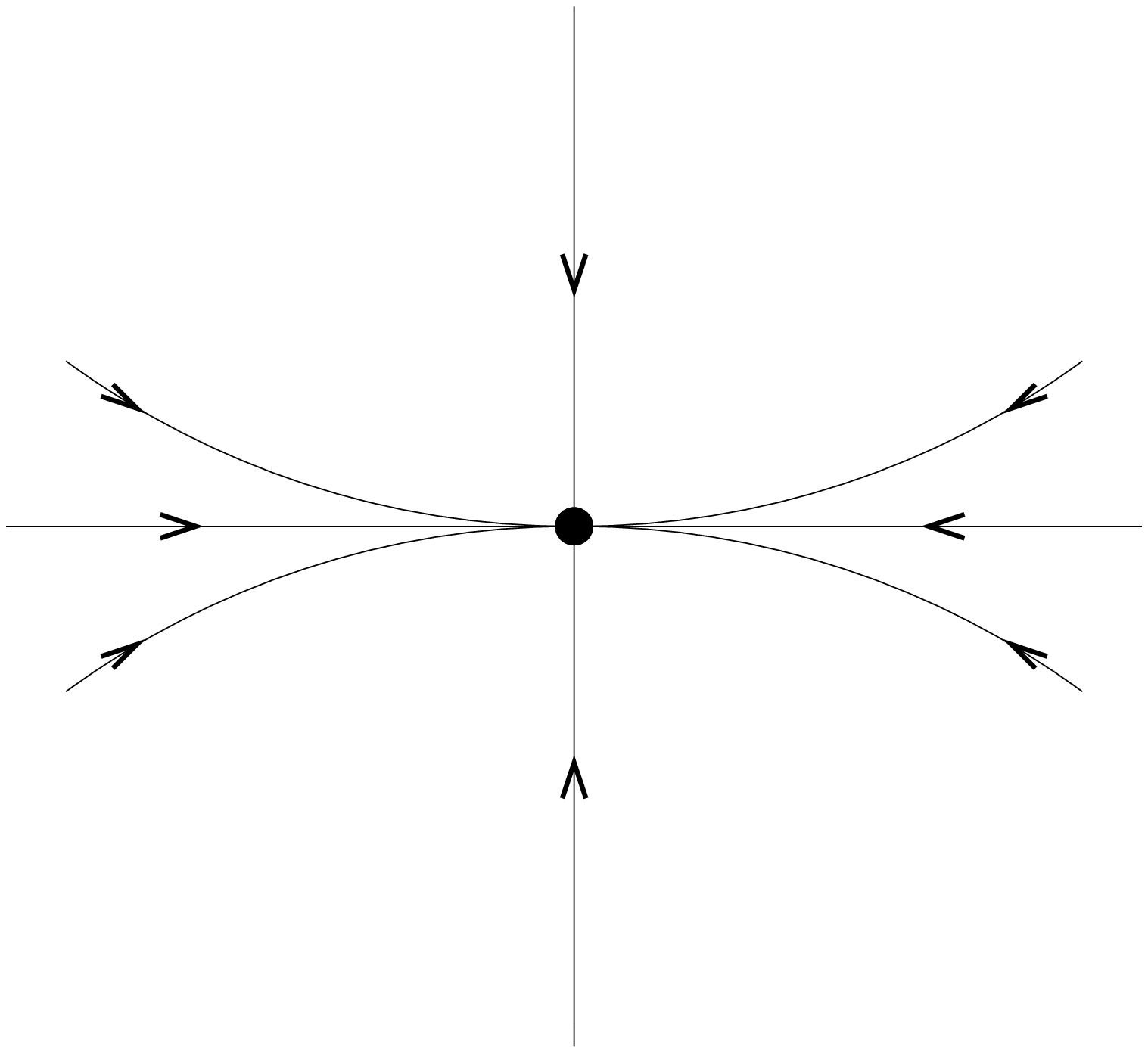,bbllx=0pt,bblly=0pt,bburx=455pt,bbury=400pt,width=7.5cm}
\psfig{file=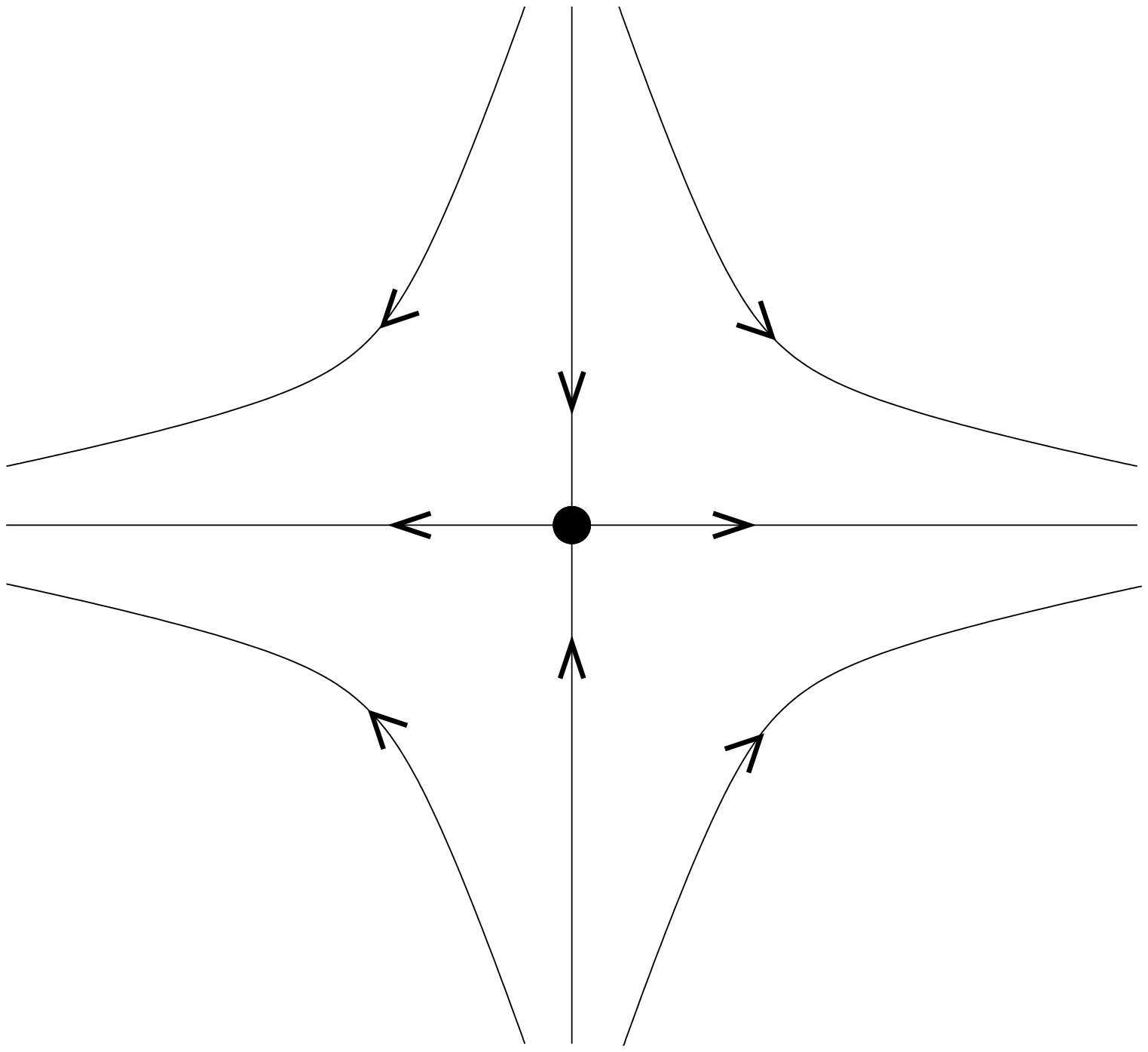,bbllx=0pt,bblly=0pt,bburx=455pt,bbury=400pt,width=7.5cm}
}}
\caption[]{Topology in the neighborhood of critical points (inspired from Jost 2002).

{\em Left panel:} expected topology of field lines nearby a
maximum, if $\lambda_1 > \lambda_2$. Except for two vertical field lines
along the eigenvector corresponding to the smallest eigenvalue, $\lambda_2$,
all the lines converging to the node tend to be aligned with the
horizontal axis, corresponding to $\lambda_1$. If $\lambda_1 =
\lambda_2$, we would face a degenerate situation where all the directions are equivalent.

{\em Right panel:} expected topology of field lines nearby a saddle
point, if $\lambda_1 > 0 > \lambda_2$. There are only four field lines
connected to this point, aligned with the two eigenvectors. The two
vertical field lines, corresponding to $\lambda_2$, are unstable since
all the other field lines are diverging away from them. Reversely, the
two horizontal lines corresponding to $\lambda_1$ are stable lines.
}
\label{figure2}
\end{figure*}

The skeleton (of over-dense regions) is defined as the borders of the
void patches (and a dual skeleton can be similarly defined as the
borders of the peak patches). It is easy to show 
that it passes through all the saddle
points and the local maxima. 
It would be out of the scope of this paper to go
further in the mathematical details of the topology of the
skeleton, but one can list the following well known 
properties, valid only if there is no unexpected degeneracies 
(e.g., Jost 2002), and which can be easily
verified by visual inspection of Fig.~\ref{figure1} (top left
panel):
\begin{itemize}
\item The nodes of the skeleton are the local maxima, where multiple
lines of the skeleton can converge. In general, because $\lambda_1 >
\lambda_2$, these lines tend to converge along the axis aligned with
the eigenvector associated to $\lambda_1$ (Fig.~\ref{figure2}, left panel). 
\item Two local maxima cannot be directly connected together, there is
always a saddle point in between;
\item Saddle points cannot be nodes of the skeleton.
Indeed, there are only four field lines connected to each saddle
point: two unstable fields lines arriving 
from opposite directions, locally parallel to the eigenvector corresponding
to $\lambda_2 < 0$, and two stable fields lines departing 
in opposite directions locally parallel to the eigenvector
corresponding to $\lambda_1 > 0$ (Fig.~\ref{figure2}, right
panel). These too last field lines locally coincide with the
skeleton and end to a local maximum.
\end{itemize}
From the last argument, the skeleton can be seen as
the ensemble of pairs of stable fields lines departing from saddle
points and connecting them to local maxima. 
The skeleton field lines can thus be drawn by going along the
trajectory with the following motion equation
\begin{equation}
\frac{d \vr}{dt}\equiv \vv=\nabla \rho, \label{eq:motion}
\end{equation}
starting from the saddle points, and with initial velocity parallel to
the major axis of the local curvature (i.e. parallel to the
eigenvector of the Hessian corresponding to $\lambda_1$).
Trajectory is followed until convergence to a local maximum.
This procedure was actually used to draw the skeleton, as explained
in details in Appendix. 

\subsection{Local approximation}
\label{sec:localapp}
As discussed in Appendix, equation of motion (\ref{eq:motion}) is not easy to
solve in practice, even for a smooth field sampled on a finite but thin
grid. Furthermore, analytic predictions are very difficult since 
eq.~(\ref{eq:motion}) is non local.
This motivates the need for an approximation of the real skeleton with a {\em
local} criterion on the density field and its derivatives of various
orders. We shall do so by two means, the first using a more mathematical
approach, the second using a more physical approach.

The mathematically motivated derivation consists in Taylor expanding the
field around its saddle points and its local maxima. On the skeleton nearby these
points, we clearly have, at leading order (and except
for degenerate cases)
\begin{equation}
\lambda_2 < 0,\label{eq:lambpos}
\end{equation}
\be
\vH \,\nabla\rho=\lambda_1 \nabla\rho. \label{eq:eigen} 
\ee
This entails a natural definition for the {\em local} skeleton: it consists
of any point of space where eqs.~(\ref{eq:lambpos}) and
(\ref{eq:eigen}) are verified.

The physically motivated approach  
consists in considering the field as a landscape, 
where the third coordinate, $r_3$, is given by $r_3=\rho(r_1,r_2)$. In
that case, the isocontour lines 
of the density field are natural objects for the analyses. 
Let us consider two pieces of isocontour lines $A$ and $B$ very close
to each other and let us move from $A$ to $B$, following the gradient.
We expect the skeleton to take either the short possible or the longest possible
path between $A$ and $B$.  Since the path length 
is inversely proportional to the magnitude of the gradient, 
the points of interest are those where $|\nabla \rho|$ is locally an
extremum along the isocontour.\footnote{An accurate examination of
the neighborhood of local maxima and saddle points suggests
that the skeleton should take the longest possible way, which 
in fact implies that the magnitude of the gradient is a local 
{\em minimum} along the contour line. However, enforcing such a condition would
lead us to examine expressions involving third derivatives
of the density field, in disagreement with a leading order approach. 
Instead, we are going to use less realistic but simpler
criteria on the local curvature of the
isocontour lines nearby the extrema of the density gradient magnitude  
to select the points of interest.}

 This translates mathematically as follows.
If we denote $s$ a curvilinear coordinate along an
isocontour, $(r_1(s),r_2(s))$, we have by definition,
\begin{equation}
\frac{\partial \rho}{\partial r_1} \frac{d r_1}{d s} +
\frac{\partial \rho}{\partial r_2} \frac{d r_2}{d s}=0, \label{eq:isocontour}
\end{equation}
with the normalization
\begin{equation}
\left(\frac{d r_1}{d s}\right)^2+\left(\frac{d r_2}{d s}\right)^2=1. \label{eq:norm}
\end{equation}
The gradient is locally an extremum along the isocontour if
\begin{equation}
\frac{d}{dt}(|\nabla \rho|^2)=0,
\end{equation}
which gives, using eq.~(\ref{eq:isocontour}),
\begin{eqnarray}
{\cal S} & \equiv & \frac{\partial \rho}{\partial r_1} \frac{\partial
\rho}{\partial r_2} \left( \frac{\partial^2 \rho}{\partial r_1^2} -
\frac{\partial^2 \rho}{\partial r_2^2} \right) \nonumber \\
& +  & \frac{\partial^2
\rho}{\partial r_1 \partial r_2}\left( \left[ \frac{\partial \rho}{\partial r_2} \right]^2 - \left[\frac{\partial
\rho}{\partial r_1} \right]^2
\right)=0.
\label{eq:sdef}
\end{eqnarray}
In fact, it is fairly easy to rewrite this equation as
\begin{equation}
{\cal S} = {\rm det}\, (\,\vH\,\nabla \rho, \nabla \rho\,)=0. \label{eq:localsketot}
\end{equation}
Therefore, the condition ${\cal S}=0$ is equivalent to say that the
gradient is an eigenvector of the Hessian. 

However, there is a supplementary condition which comes out  naturally:
while walking from one field line to another, one prefers
to stay on a ridge, that is on the points where the curvature of the
isocontour is positive, i.e.
\begin{equation}
{\cal C}\equiv \frac{\nabla \rho}{|\nabla \rho|}\cdot \frac{d^2 \vr}{d s^2} >
0,
\end{equation}
which translates in, after some algebra based on
eqs.~(\ref{eq:isocontour}) and (\ref{eq:norm}), 
\begin{equation}
{\cal C}=-\frac{1}{|\nabla \rho|^3}\, {}^{\rm t}{\nabla \rho_{\perp}} \,
\vH \, {\nabla \rho_{\perp}} > 0, \label{eq:condcur}
\end{equation} 
where $\nabla \rho_{\perp} \equiv (\partial \rho/\partial r_2, -\partial
\rho/\partial r_1)$. 
So, we have to select among the points verifying
eq.~(\ref{eq:localsketot}) those which have ${\cal C} > 0$.
Since, $\nabla\rho$ is an eigenvector of $\vH$, so is $\nabla
\rho_{\perp}$, therefore
\begin{equation}
{\cal C}=-\frac{\lambda_2}{|\nabla \rho|} \ {\rm or} \
-\frac{\lambda_1}{|\nabla \rho|}.
\end{equation}
After a simple examination of the various cases, 
$\lambda_1 \geq \lambda_2 > 0$, $\lambda_1 > 0 > \lambda_2$ and $0 >
\lambda_1 \geq \lambda_2$, we finally obviously converge again to 
eqs.~(\ref{eq:lambpos}) and (\ref{eq:eigen}), except for hills, $0 >
\lambda_1 \geq \lambda_2$, where equation (\ref{eq:condcur}) allows
the gradient to be aligned with both axes of the curvature. In this
last case, we can see however that the situation 
$\vH \,\nabla\rho=\lambda_2 \nabla\rho$ contradicts the ``natural''
definition of a ridge. Such a ridge would indeed be more curved
along its path than orthogonally to it, a situation clearly
unrealistic in the neighborhood of a local maximum.

\subsection{Examples}
\label{sec:illustration}

\begin{figure*}
\centerline{\hbox{
\psfig{file=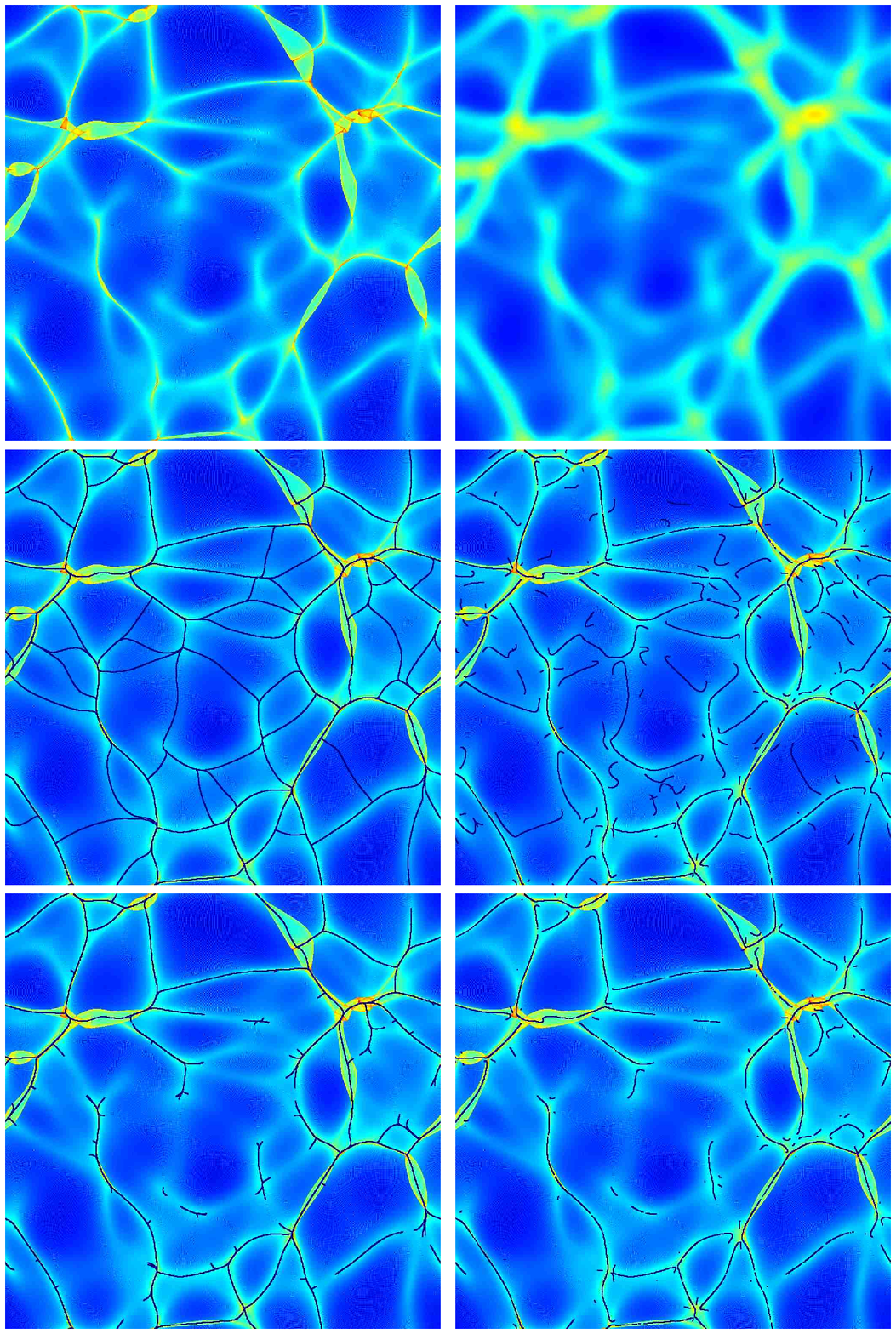,bbllx=116pt,bblly=159pt,bburx=523pt,bbury=763pt,width=12.5cm}
}}
\caption[]{Skeleton and its local approximation for the Zel'dovich
mapping of the smoothed Gaussian field of Fig.~\ref{figure0}. 

{\em Upper left panel:} Zel'dovich mapping of the smoothed field: the
Lagrangian displacement field $\vP$ was normalized so that $\nabla.\vP =
-(\rho-\langle \rho \rangle)/\sigma \langle \rho \rangle$, where
$\sigma^2$ is the variance of the initial smooth map.

{\em Upper right panel:} the field of the top left panel smoothed
with a Gaussian window of radius 12.5 pixels. Such smoothing is
necessary to get rid of caustics and to enforce the differentiability
required to measure the skeleton. The smoothing scale is such that the large scale
features outside caustics are preserved: it has to be small compared to the initial
smoothing radius of 25 pixels and large enough compared to the pixel size.

{\em Middle left and middle right panels:} respectively, the real and the local
skeleton superposed to the Zel'dovich map. 

{\em Lower left and lower right panels:} respectively, the real and the local
skeleton superposed to the Zel'dovich map, but restricted to over-dense regions. 
}
\label{figure3}
\end{figure*}

We now compare the local and real skeleton by visual inspection of
Figs.~\ref{figure1} and~\ref{figure3}, which respectively correspond
to a Gaussian realization and its Zel'dovich mapping. We also measure
the length of the skeletons as a function of threshold for these two
particular examples. 

Clearly the local skeleton is an excellent approximation of the real
one. Indeed most of the large scale features are very well
captured, particularly in the vicinity of maxima and saddle points,  
as a result of our perturbative approach. 
In agreement with intuition, the more filamentary is the
field, the better is the agreement: the local skeleton seems to
perform better in the Zel'dovich map than in the Gaussian
one. However, connectivity of the local skeleton is not ensured, at
variance with the real one. Furthermore, there are little spurious
structures in void  patches that do not match any line of the real skeleton. By
restricting the comparison to over-dense regions, a large number of
these structure disappear and the agreement improves significantly, at
least visually.

\begin{figure*}
\centerline{\hbox{
\psfig{file=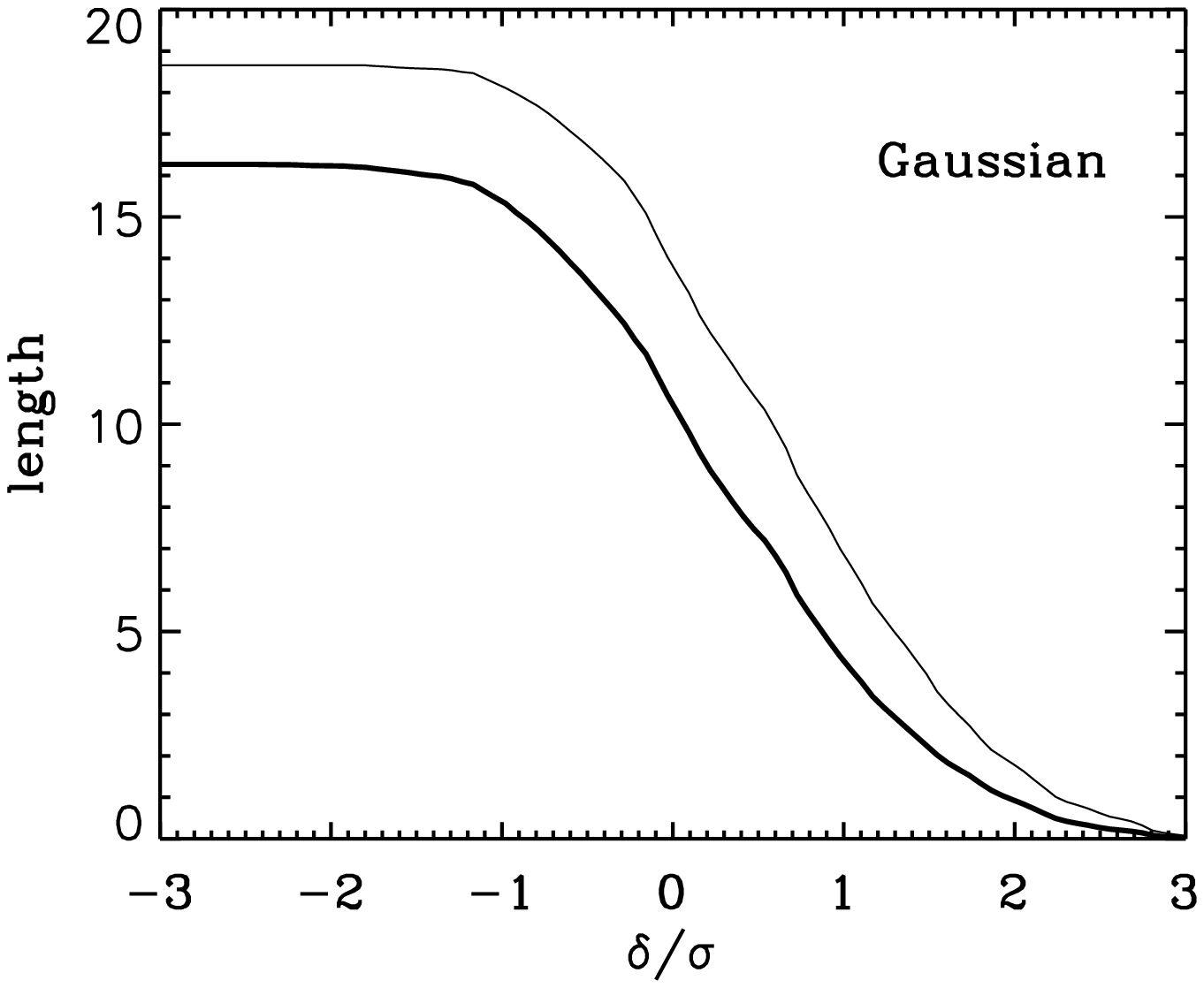,bbllx=132pt,bblly=363pt,bburx=542pt,bbury=699pt,width=7cm}
\psfig{file=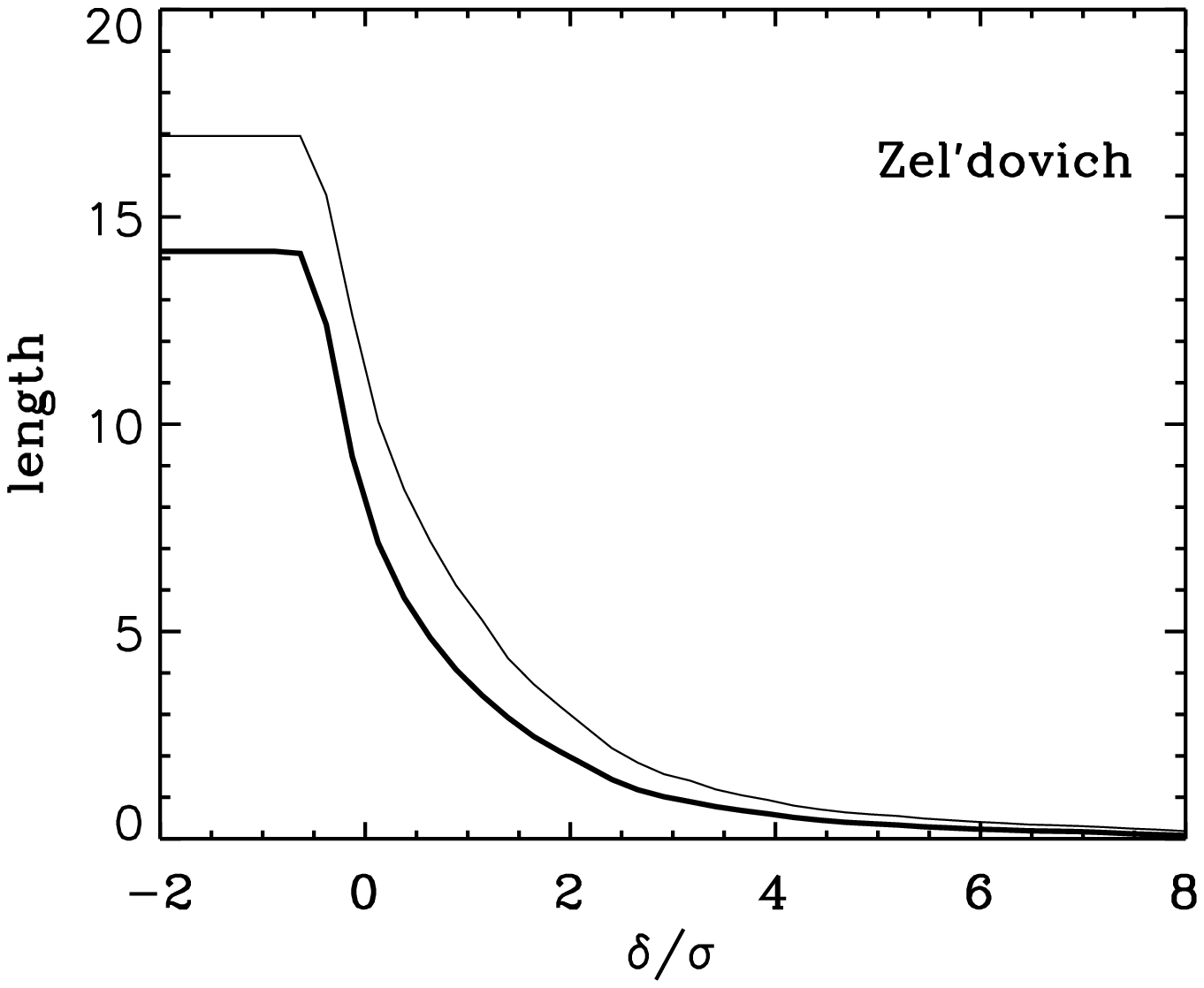,bbllx=132pt,bblly=363pt,bburx=542pt,bbury=699pt,width=7cm}
}}
\caption[]{Real versus local skeleton length comparison. 

{\em Left panel:} the skeleton length in units of sample box length
as a function of density
threshold, measured for the real (thin curve) and the local (thick
curve) skeleton in the smooth Gaussian field of Fig.~\ref{figure0}.
$\delta/\sigma\equiv (\rho-\langle \rho \rangle)/\sigma \langle \rho
\rangle$ is the density contrast in units of the variance of the
smoothed field.

{\em Right panel:} same as left panel but for the smoothed Zel'dovich
map (top right panel of Fig.~\ref{figure3}). }
\label{figure4}
\end{figure*}

A more quantitative analysis can be conducted by comparing the
measured length of the real and local skeletons
in regions where the density exceeds a given threshold, as illustrated
by Fig.~\ref{figure4}. Contrary to what would suggest the visual
inspection of Figs.~\ref{figure1} and \ref{figure3}, 
the local skeleton is systematically slightly shorter than the
real one. The total lengths differ by about  20
percents, both for the Gaussian smooth field and its Zel'dovich
mapping. However, as illustrated by left panel of Fig.~\ref{figure2},
skeleton field lines converging to a local maximum tend to superpose
along the major axis of the local curvature, which produces multiple
lines. This feature inherent to the Lagrangian nature of the real
skeleton is missing in the local skeleton, due to its local, 
Eulerian nature (see also Appendix). Hence, it is not surprising that
the local skeleton is shorter than the real one. 
 
\section{Link to statistics}
\label{sec:stat}
In this section we focus on the local skeleton. For the sake of
simplicity, we study from now on the {\em total} local skeleton, defined
{\em as the full set of points satisfying the condition} ${\cal S} = 0$
[eq.~(\ref{eq:localsketot})]. We first examine 
the Gaussian case in \S~\ref{sec:gaussian}, where 
specific analytic results are derived
and then confronted to numerical experiments. 
The normalized differential length of the skeleton 
as a function of density threshold is seen to be 
very close to the probability distribution function (pdf) 
of the smoothed field, i.e. a Gaussian.

We thus consider in section \S~\ref{sec:ngaussian} some examples of non-Gaussian
fields, namely $\chi^2$ distributions with $n$ degrees of freedom, the
Zel'dovich mapping discussed previously and, finally, an extreme case
where the density contrast is locally enhanced along lines with random
orientations. In all these cases, we find again that the differential
length of the skeleton scales very much like the pdf, with slightly
worse noise properties as expected since this estimator relies on
derivatives of the field. This intriguing result, despite its
mathematical beauty, might look discouraging for using the skeleton as
a test of non-gaussianity. However, our analysis does not use the
supplementary information provided by the skeleton, namely its
total length, which is considered here as an arbitrary normalization. 

\subsection{The Gaussian case}
\label{sec:gaussian}

In this section, 
after defining a set of useful notations (\S\ref{sec:spec}), we
compute a general expression for the differential length of
the skeleton, $f$, as a function of density threshold (\S\ref{sec:ske}).
Then we concentrate on the Gaussian case (\S\ref{sec:anal}) 
and show that the shape of $f$  depends only on a single spectral parameter, 
$\gamma$, defined below. We demonstrate that for $\gamma=0$, $f$ is exactly
given by a Gaussian. Given the level of
complexity of the calculations, we are however lead to rely on
a power expansion in $\gamma$ around a Gaussian 
to examine the case $\gamma > 0$.
This power expansion is checked carefully 
against numerical experiments (\S\ref{sec:gaus}), which indeed
show that $f$ deviates only weakly from a Gaussian. 

\subsubsection{Spectral parameters and dimensionless variables}
\label{sec:spec}
From now on, we assume without loss of generality that the random 
smoothed field, $\rho$ has zero average $\langle \rho \rangle=0$.
For convenience we define the following spectral parameters:
\bea
\vspace{0.2cm}
\sigma^2_0 & = & \langle\rho^2\rangle,\\
\vspace{0.2cm}
\sigma^2_1 & = & 2\langle\rho^2_1\rangle=2\langle\rho^2_2\rangle,\\
\vspace{0.2cm}
\sigma^2_2 & = & \frac{8}{3}\langle\rho^2_{11}\rangle=
\frac{8}{3}\langle\rho^2_{22}\rangle=8\langle\rho^2_{22}\rangle,\\
\sigma^2_3 & = & \frac{16}{5}\langle\rho^2_{111}\rangle=
\frac{16}{5}\langle\rho^2_{222}\rangle=
16\langle\rho^2_{112}\rangle=16\langle\rho^2_{122}\rangle,\\
\gamma     & = & \sigma_1^2/(\sigma_0\sigma_2),\\
\widetilde{\gamma} & = & \sigma_2^2/(\sigma_1\sigma_3),
\eea
where  $\rho_i\equiv \partial \rho/\partial r_i$, $\rho_{ij}\equiv 
\partial^2 \rho/\partial r_i \partial r_j$ and
$\rho_{ijk}\equiv \partial^3 \rho/\partial r_i \partial r_j \partial r_k$ ($i,j,k=1,2$)
are its gradient, hessian and matrix of third derivatives, respectively. 
Using these parameters one can consider the 10 following dimensionless variables: 
\begin{equation}
x=\frac{\rho}{\sigma_0},\hspace{0.5cm} 
x_i=\frac{\rho_i}{\sigma_1},\hspace{0.5cm} 
x_{ij}=\frac{\rho_{ij}}{\sigma_2},\hspace{0.5cm} 
x_{ijk}=\frac{\rho_{ijk}}{\sigma_3}, 
\end{equation}
and the dimensionless function $s = {\cal  S}/(\sigma_1^2\sigma_2)$. 
That is, according to eq. (\ref{eq:sdef}):
\begin{equation}
s = x_1x_2(x_{11}-x_{22})+x_{12}(x_2^2-x_1^2).\label{eq:lsdef}
\end{equation}
Thus, the points of the random field where the first and second derivatives satisfy 
the condition $s=0$ define the total local skeleton. Finally, the derivatives 
\be
s_i = {\cal S}_i/(\sigma_1^2\sigma_3)
\en 
will also be useful. It is worth noting here that $s_i$ depends on $\widetilde{\gamma}$,
but not on $\gamma$, i.e. $s_i=s_i(x_i,x_{ij},x_{ijk},\widetilde{\gamma})$.

\subsubsection{Length of the skeleton: general expressions}
\label{sec:ske}
If we denote $\P_s(\rho,\cS,\cS_1,\cS_2)$ 
the joint probability distribution of the variables $\rho$,
$\cS$, and $\cS_i$ for $i=1,2$, the expected average of the 
skeleton length $\L(\rho_{\rm th})$ per unit area 
above some threshold $\rho_{\rm th}$ can be derived the following way. 

Let us consider a straight line along the direction $r_1$ and 
which intersects the isocontour lines, $\cS=0$, at some point, where  
$\rho>\rho_{\rm th}$. In the vicinity of such a point, where $\cS=0$ and
$d\cS=\cS_1dr_1$, we can integrate $\P_sd\rho d\cS d\cS_1 d\cS_2$ over $dr_1$
from $-dr_1/2$ to $dr_1/2$ and we get, obviously:
\bea
\int\limits_{r_1 \in [-dr_1/2,dr_1/2],\  \rho > \rho_{\rm th}} 
d\cS d\cS_1d\cS_2d\rho\P_s(\rho,0,\cS_1,\cS_2)
=\nonumber\\
dr_1\int\limits_{\rho > \rho_{\rm th}}|\cS_1|d\cS_1d\cS_2
d\rho\P_s(\rho,0,S_1,S_2).\label{eq:sint}
\ena
This integral represents the probability to find the point 
$\cS=0$ along the line $r_2=$const in the
range $[r_1-dr_1/2,r_1+dr_1/2]$. Note that the absolute value 
of $\cS_1$ is here necessary since we want to take into account 
both up-crossing and down-crossing points. Then, 
the elementary length of the isocontour line $\cS=0$ inside the square 
$[r_1-dr_1/2,r_1+dr_1/2;r_2-dr_2/2,r_2+dr_2/2]$ is $dr_2/\cos(\alpha)$,
where $\alpha$ is the angle between $r_2$ and the isocontour line. Since 
$\cos(\alpha)=|\cS_1|/\sqrt{\cS_1^2+\cS_2^2}$, using eq.~(\ref{eq:sint}) 
one gets 
\bea
\lefteqn{\L(\rho_{\rm th})dr_1dr_2 = }   \nonumber \\
&&dr_1dr_2\int\limits_{\rho > \rho_{\rm th}} d\rho
d\cS_1d\cS_2\sqrt{\cS_1^2+\cS_2^2}\ \P_s(\rho,0,\cS_1,\cS_2).
\ena
This equation represents the average length of the skeleton per element of area $dr_1dr_2$. 
In terms of dimensionless variables, it rewrites
\be
\L(x_{\rm th})=\int\limits_{x> x_{\rm th}} dx
ds_1 ds_2\ \frac{\sigma_3}{\sigma_2}\sqrt{s_1^2+s_2^2}\ \P_s(x,0,s_1,s_2).\label{eq:ldef}
\en
We shall now look for an analytical expression for this function $\L$ in the case 
of a Gaussian field.
\subsubsection{Towards an analytic expression for a Gaussian field}
\label{sec:anal}
Deriving eq.~(\ref{eq:ldef}) we did not consider any special features
of the joint probability function $\P_s$, thus this equation is true
for any random field. Unfortunately the derivation of function $\P_s$ is
not easy, even in the Gaussian case, that we examine now. 

We consider the 10 components random vector, $\a$,  
\be
\a = (x,x_i,x_{ij},x_{ijk})\quad  (i,j,k=1,2),
\en
and the probability distribution function $\P(\a)$. In the Gaussian
case, this can be written as
\be
\P(\a) = \frac{1}{[(2\pi)^{10}\, |\M|]^{-\frac{1}{2}}}\ \exp\left[{-{1\over2}\a\M^{-1}\a^T}\right]
\en
where $\M$ is the covariance matrix, $\M=\langle \a\a^T \rangle$, and $|\M|$ its
determinant. Then,
eq.~(\ref{eq:ldef}) can be rewritten using $\P(\a)$
\be
\L(x_{\rm th}) =
\int\limits_{x>x_{\rm th},\ s=0}\frac{\sigma_3}{\sigma_2}\sqrt{s_1^2+s_2^2}\P(\a)d\a.
\label{eq:jenesaispas}
\en
For further investigation, it is particularly convenient to consider the following variables:
\be
q
=x_{11}-x_{22},\quad u=2x_{12},\quad v=x_{11}+x_{22}. \label{eq:qdef}
\en
We are interested in the distribution of the skeleton length over just
one variable $x_{\rm th}$. It is worth mentioning that using the variables
defined in eq.~(\ref{eq:qdef}), $x$ correlates only with $v$ and
does not correlate with either $q$, $u$, $x_i$ nor $x_{ijk}$, i.e. 
\bea
\langle xq \rangle & = & \langle xu\rangle=\langle xx_i\rangle=\langle xx_{ijk}\rangle=0,\\
\langle xv\rangle & = & -\gamma. \label{eq:qdef2}
\ena
Taking into account eqs.~(\ref{eq:qdef}) and (\ref{eq:qdef2}), 
one can represent $\P(\a)$ in the following way:
\be
\P(\a) = \frac{1}{\sqrt{1-\gamma^2}}\exp\left[{-\frac{(x+\gamma v)^2}{2(1-\gamma^2)}}\right]\ 
\widetilde{\P}(x_i,q,u,v,x_{ijk}).
\en
In order to take into account the condition $s=0$ in eq.~(\ref{eq:jenesaispas}), 
we should perform one more substitution:
\bea
x_1 = r\cos(\varphi), &\quad & x_2 = r\sin(\varphi), \\
q   = p\cos(2\psi),   &\quad & u   = p\sin(2\psi).
\ena
It is easy to see that with these new variables, the function $s$ in
eq.~(\ref{eq:lsdef}) reads:
\begin{equation}
s=\frac{1}{2}pr^2\sin\bigl[2(\varphi-\psi)\bigr].
\end{equation}
The condition $s=0$ is now transformed into a simple relation between
the angles $\varphi$ and $\psi$.  
The length of the skeleton can therefore be written as follows:
\bea
\lefteqn{\L(x_{\rm th}) = \frac{\sigma_3}{\sigma_2}
\int\limits_{x_{\rm th}}^{\infty}dx\int\limits_{-\infty}^{\infty}dv
\frac{e^{-\frac{(x+\gamma v)^2}{2(1-\gamma^2)}}}{\sqrt{1-\gamma^2}}\int
d\Omega\ \sqrt{s_1^2+s_2^2}\ \times}                                \nonumber\\
& \bigl[\delta_{\rm D}(\varphi-\psi)+\delta_{\rm D}(\varphi-\psi+\pi/2)\bigr]\ 
{\hat\P}(r,p,\varphi,\psi,x_{ijk},v)
\label{eq:lgauss}
\ena
where $d\Omega=p\,dp\,r\,dr\,dx_{ijk}\,d\psi\,d\varphi$ and $\delta_{\rm D}$ is the
usual delta function. This equation will give us the total length of
the skeleton $\L_{\rm tot}$ if we consider $x_{\rm th} \rightarrow
-\infty$. From eqs.~(\ref{eq:lsdef}) and (\ref{eq:lgauss}) one can see
that the differential length normalized by the total length is: 
\be
f \equiv -\frac{1}{\L_{tot}}\frac{\partial \L}{\partial x} = 
\int\limits_{-\infty}^{+\infty} dv\ {\cal C}(v)\ 
\frac{e^{-\frac{(x+\gamma v)^2}{2(1-\gamma^2)}}}{\sqrt{1-\gamma^2}}.\label{eq:fdef}
\en
Therefore, remarkably, $f$ is a function of $x$ and only one spectral parameter $\gamma$.
The quantity $f(x,\gamma)dx$ simply represents the fraction of the skeleton length
between the levels $x$ and $x+dx$.  However, the unknown function ${\cal
  C}(v)$ is rather cumbersome to estimate analytically. 
We examine in next section a way to avoid 
its calculation, but which relies partly on numerical experiments.

\subsubsection{Final expression in the case of a Gaussian field}
\label{sec:gaus}
The first thing to notice, when examining eq.~(\ref{eq:fdef}), is that 
in the limit $\gamma=0$, $f$ is exactly a Gaussian:
\be
f(x,\gamma=0)=\frac{1}{\sqrt{2\pi}} e^{-x^2/2}.
\ee
We thus
expect $f$ to depart only weakly from a Gaussian if $\gamma$ is small
enough, which motivates for a power-expansion of $f(x,\gamma)$ in $\gamma$
around a Gaussian.

According to eq.~(\ref{eq:fdef}), $f(x,\gamma)$ satisfies the
following partial differential equation:
\be
\frac{\partial f}{\partial\gamma}\gamma=-\frac{\partial}{\partial x}
\left(\frac{\partial f}{\partial x}+xf\right).\label{eq:fdif}
\en
Since $f(x,\gamma)$ is a probability distribution function, it should
satisfy the following condition:
\be
\int f(x,\gamma)dx=1. \label{eq:fint}
\en
To try to solve eq.~(\ref{eq:fdif}), we examine solutions of the form
\be
f(x,\gamma)=\left\{ \sum_{n \geq 0} g_n(x) \gamma^n \right\} \frac{1}{\sqrt{2\pi}} e^{-x^2/2}.
\ee 
Injecting this expression in eq.~(\ref{eq:fint}) leads to
\be
\frac{d^2g_n}{dx^2}-x \frac{dg_n}{dx} + n g_n=0.
\label{eq:Herm1}
\ee
Setting $y=x/\sqrt{2}$, we find
\be
\frac{d^2g_n}{dy^2}-2 y \frac{dg_n}{dy}+2 n g_n=0,
\ee
a differential equation followed by Hermite polynomials, $H_n(y)$.
A more detailed examination of the possible solutions
gives
\be
g_n(x)=H_n(x)\left[ C_n + D_n \int^{x/\sqrt{2}} dy\ \exp(y^2) /H^2_n(y) \right].
\ee
Questionable arguments based on enforcing 
the convergence of the moments of $f$ with respect to $x$ suggest
$D_n=0$. 
Symmetry of the Gaussian field with respect to $x=0$ implies $C_{2n+1}=0$.
As a result we expect $f$ to have the following form
\be
f(x,\gamma)=\left\{ \sum_{n \geq 0} C_{2n} H_{2n}(x/\sqrt{2}) 
\gamma^{2n} \right\} \frac{1}{\sqrt{2\pi}} e^{-x^2/2}.
\label{eq:edge}
\ee
with $C_0=1$, from normalization (\ref{eq:fint}).
Note that this expression is nothing but a Gram-Charlier expansion prior
to standardization [see Stuart \& Ord 1994, eq.~(6.32)].
It can be valid in practice only if the departure from a Gaussian is weak. Alternatively,
thus, we could have derived eq.~(\ref{eq:edge}) by trying to find
solutions of the form
\be
f(x,\gamma)=\left\{ \sum_{n \geq 0} h_{2n}(\gamma) H_{2n}(x/\sqrt{2}) 
\right\} \frac{1}{\sqrt{2\pi}} e^{-x^2/2},
\ee
using directly the Gram-Charlier expansion. 

\begin{figure*}
\centerline{
\hbox{
\psfig{file=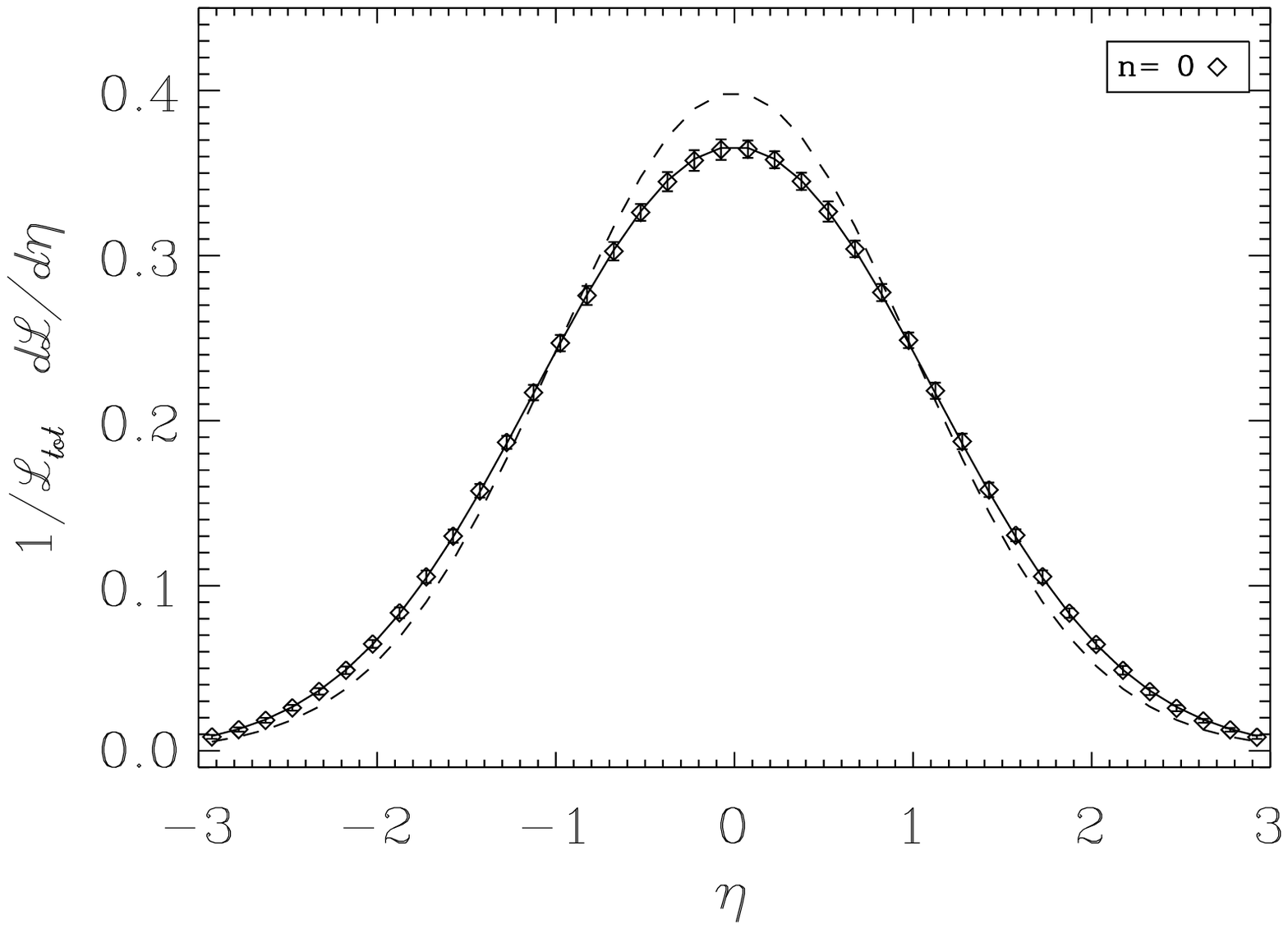,bbllx=66pt,bblly=364pt,bburx=539pt,bbury=700pt,width=7.0cm}
\psfig{file=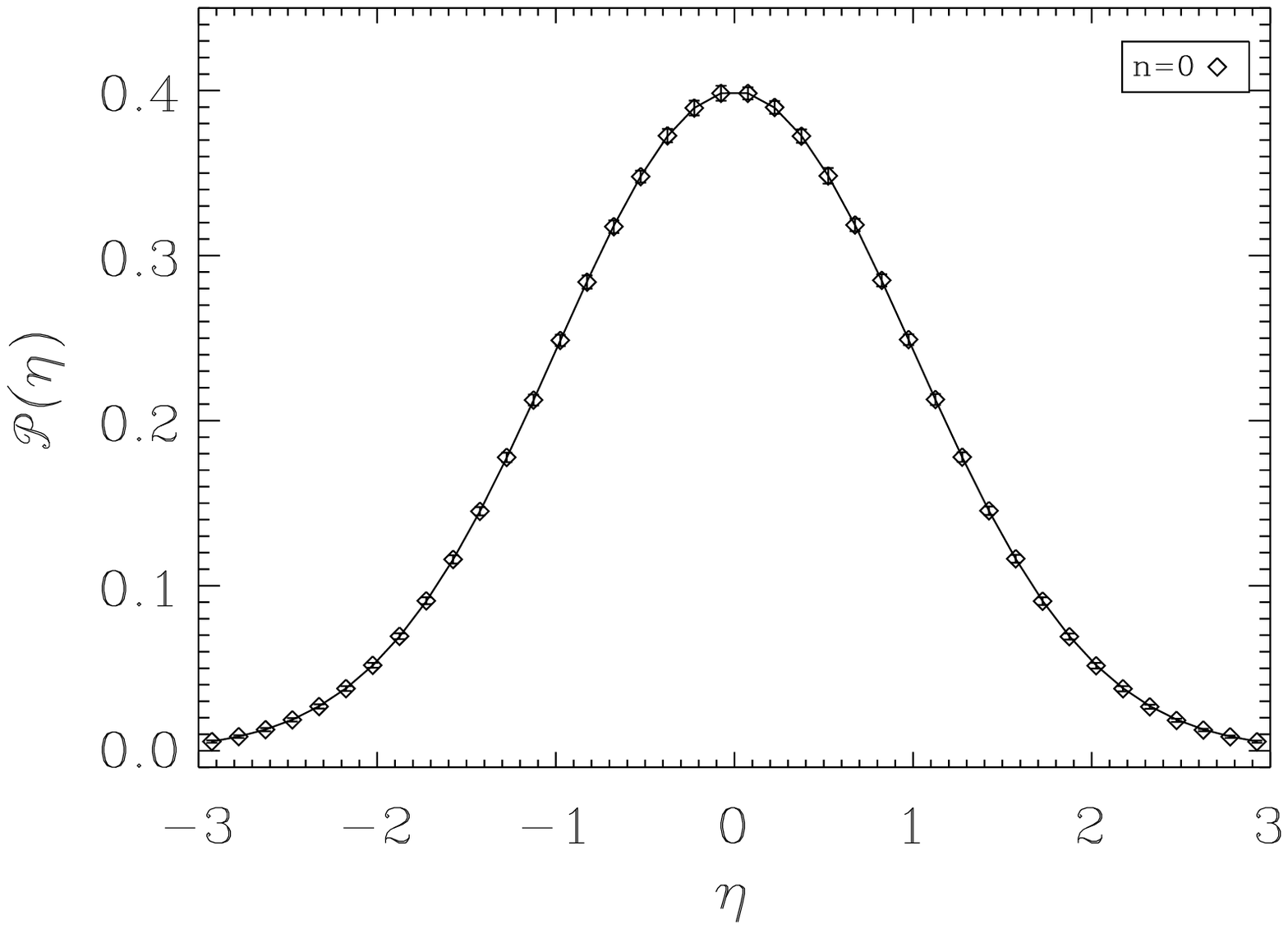,bbllx=66pt,bblly=364pt,bburx=539pt,bbury=700pt,width=7.0cm}
}}
\centerline{
\hbox{
\psfig{file=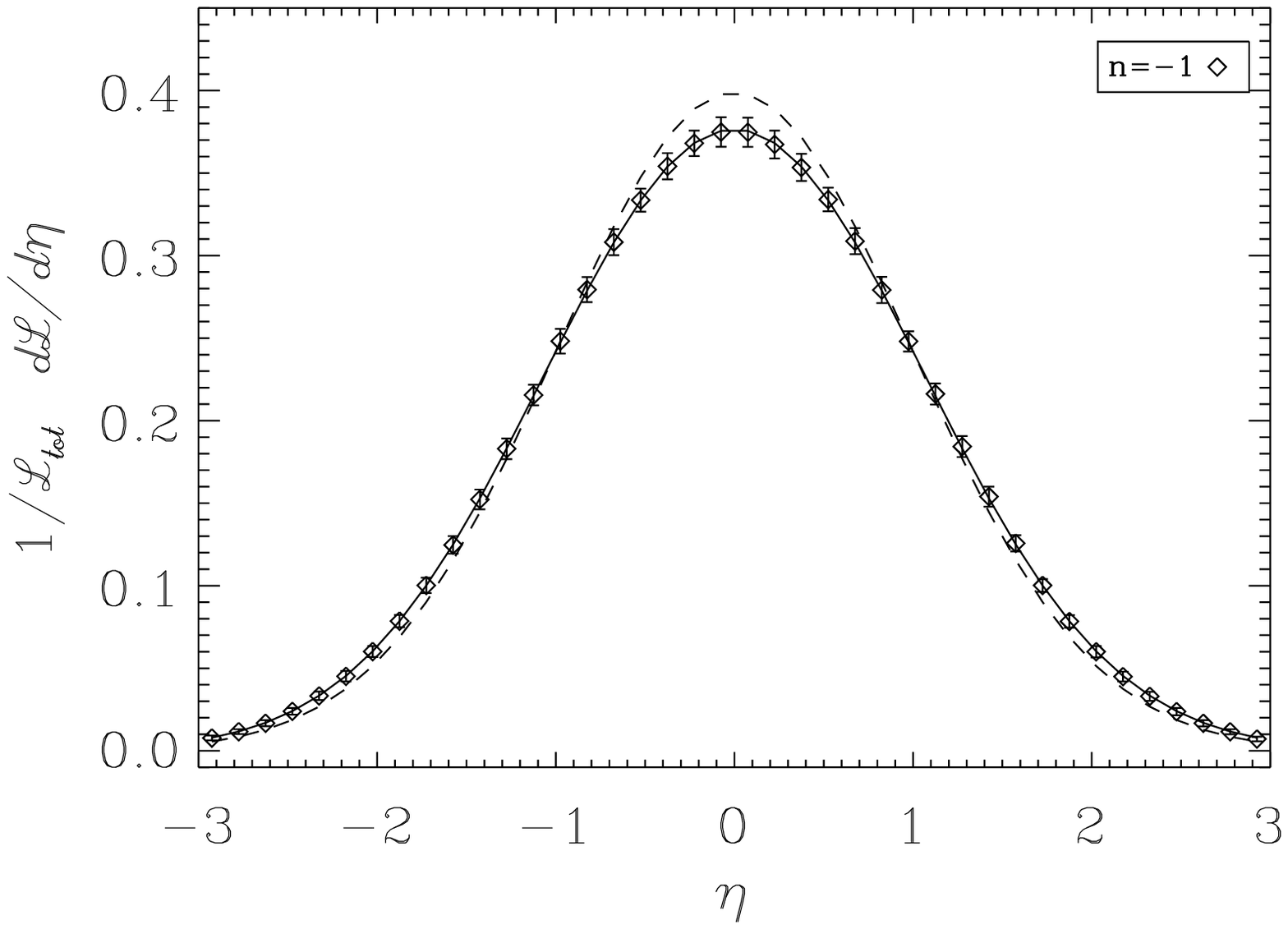,bbllx=66pt,bblly=364pt,bburx=539pt,bbury=700pt,width=7.0cm}
\psfig{file=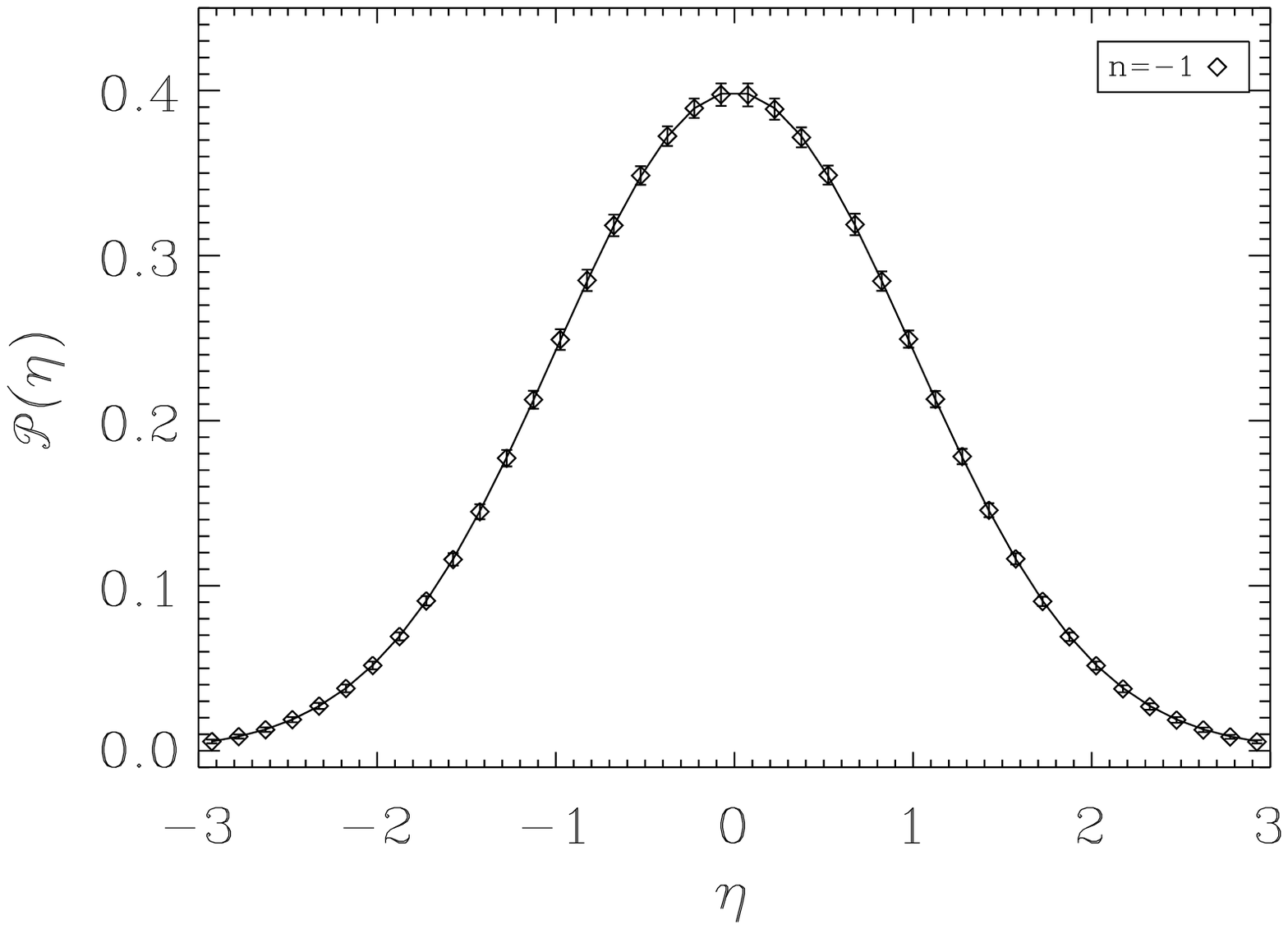,bbllx=66pt,bblly=364pt,bburx=539pt,bbury=700pt,width=7.0cm}
}} 
\centerline{
\hbox{
\psfig{file=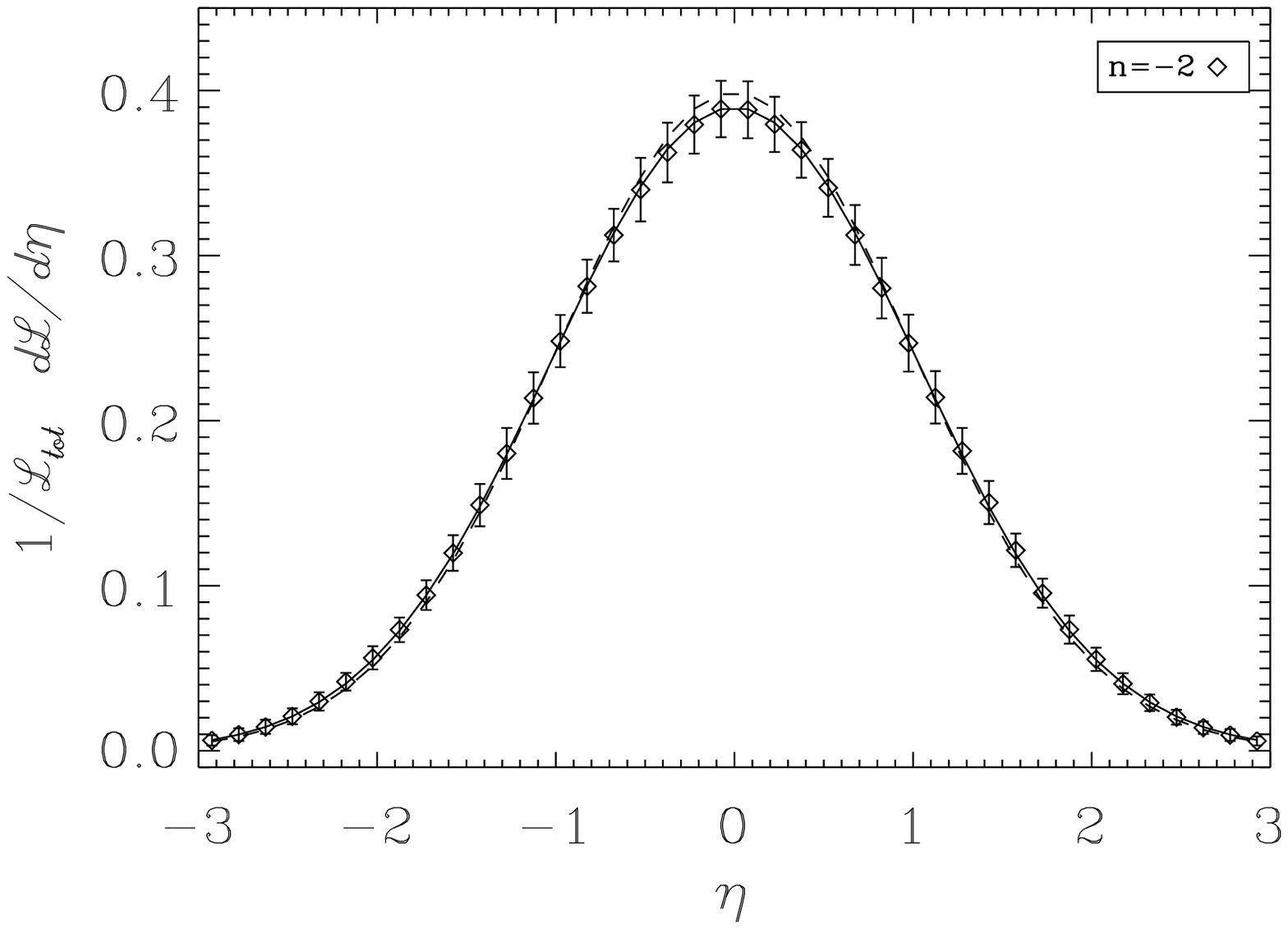,bbllx=66pt,bblly=364pt,bburx=539pt,bbury=700pt,width=7.0cm}
\psfig{file=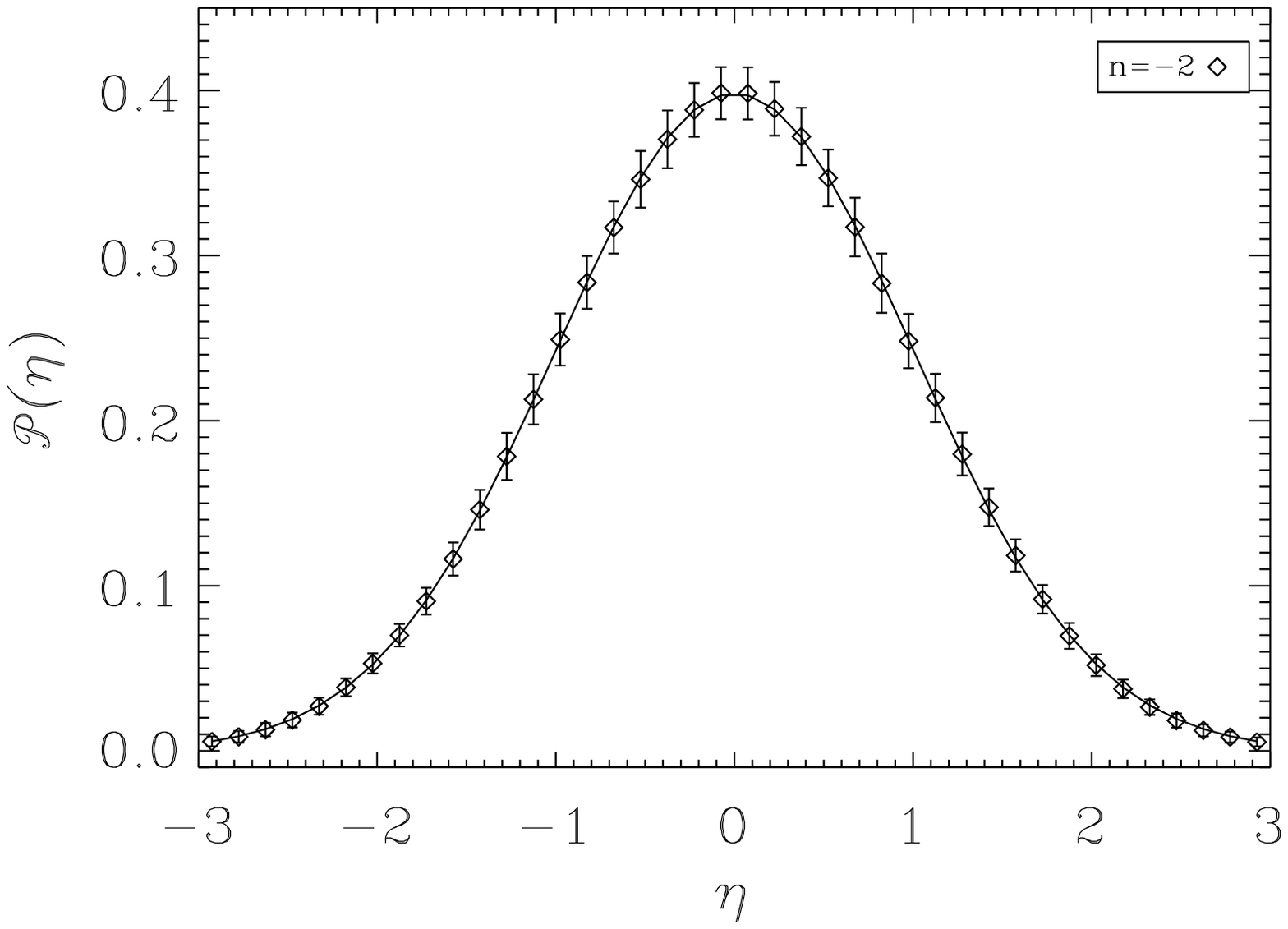,bbllx=66pt,bblly=364pt,bburx=539pt,bbury=700pt,width=7.0cm}
}}
\caption[]{The measured differential length of the total local skeleton 
[eq.~(\ref{eq:localsketot}), left panels] and the
pdf of the smoothed field for scale-free Gaussian random fields (right panels) with
power-spectra $P(k) \propto k^n$, $n=0$, $-1$ and $-2$ as indicated
on each panel, as functions
of normalized density contrast, $\eta=\delta/\sigma=x=\rho/\sigma$. 
For each value of $n$, 100 realizations were performed on a periodic grid of
size $1024\times1024$ pixels ($2048\times 2048$ for $n=0$), 
and then smoothed with a Gaussian window
of radius 5 pixels (10 pixels for $n=0$). On left panels, the dashed and solid curves 
correspond respectively to the Gaussian
limit and our semi-analytic expression (\ref{eq:fgauss}).
On right panels, the solid curves correspond to the Gaussian limit.
  The symbols with errorbars are the measurements.}
\label{figure8}
\end{figure*}
The coefficients $C_{2n}$, $n \geq 1$, remain to be determined.
We therefore performed a set of numerical
experiments which was used as well to test extensively our
skeleton analysis software (see discussion in Appendix). 
We generated scale-free random Gaussian
fields with power-spectra $P(k) \propto k^n$, $n=0$, $-1$, and
$-2$ ($\gamma=0.71,\,0.58,\,0.32$, respectively).
For each value of $n$ we performed 100 realizations over a periodic
grid of size $1024\times1024$ pixels (except for $n=0$,
where we used $2048\times2048$ pixels).
The field was smoothed with
a window of radius 5 pixels (10 pixels for $n$=0).\footnote{The case $n=0$
requires a larger smoothing window, compared to the pixel size, see discussion in 
Appendix. 10 pixels is a rather conservative but safe choice for $n=0$. It implies
to use images of $2048\times2048$ pixels,
in order to preserve the ratio between the size of the smoothing window and
the total size of the image, chosen for all values of $n$ to be approximately
equal to $1/200$.}  
The results are shown in
Figure~\ref{figure8} for the differential skeleton length 
(left panels) and for the
measured pdf (right panels).  The errorbars are obtained by the
scatter over the 100 realizations. They are of the same order 
for the skeleton as for the pdf, although slightly larger for the
former than for the latter. These small differences will be more
visible and explained in the next section, 
which deals with non Gaussian cases.

We see that the departure of $f(x,\gamma)$ from a Gaussian (dotted curve on
all the panels) is quite
weak, even for $n=0$, which corresponds to a large value of 
$\gamma=0.71$. As a result, only first order correction is needed, and
we find numerically that
\bea
C_0    & = & 1, \quad C_2=0.17,\\
C_{2n} & = & 0\quad \textrm{for}\quad n>2,
\eea
provides an excellent approximation to $f(x,\gamma)$ in the Gaussian
limit (solid curve on each left panel).
Our final expression for the normalized differential
length of the total local skeleton, is thus, for a smoothed Gaussian field:
\be
\label{eq:fgauss}
f(x,\gamma)=\frac{1}{\L_{tot}}\frac{\partial \L}{\partial x}\simeq
\frac{1}{\sqrt{2\pi}}e^{-x^2/2}\left[ 1+0.17\ \gamma^2(1-x^2)\right].
\en

\subsection{The non Gaussian case}
\label{sec:ngaussian}
\begin{figure*}
\centerline{\hbox{
\psfig{file=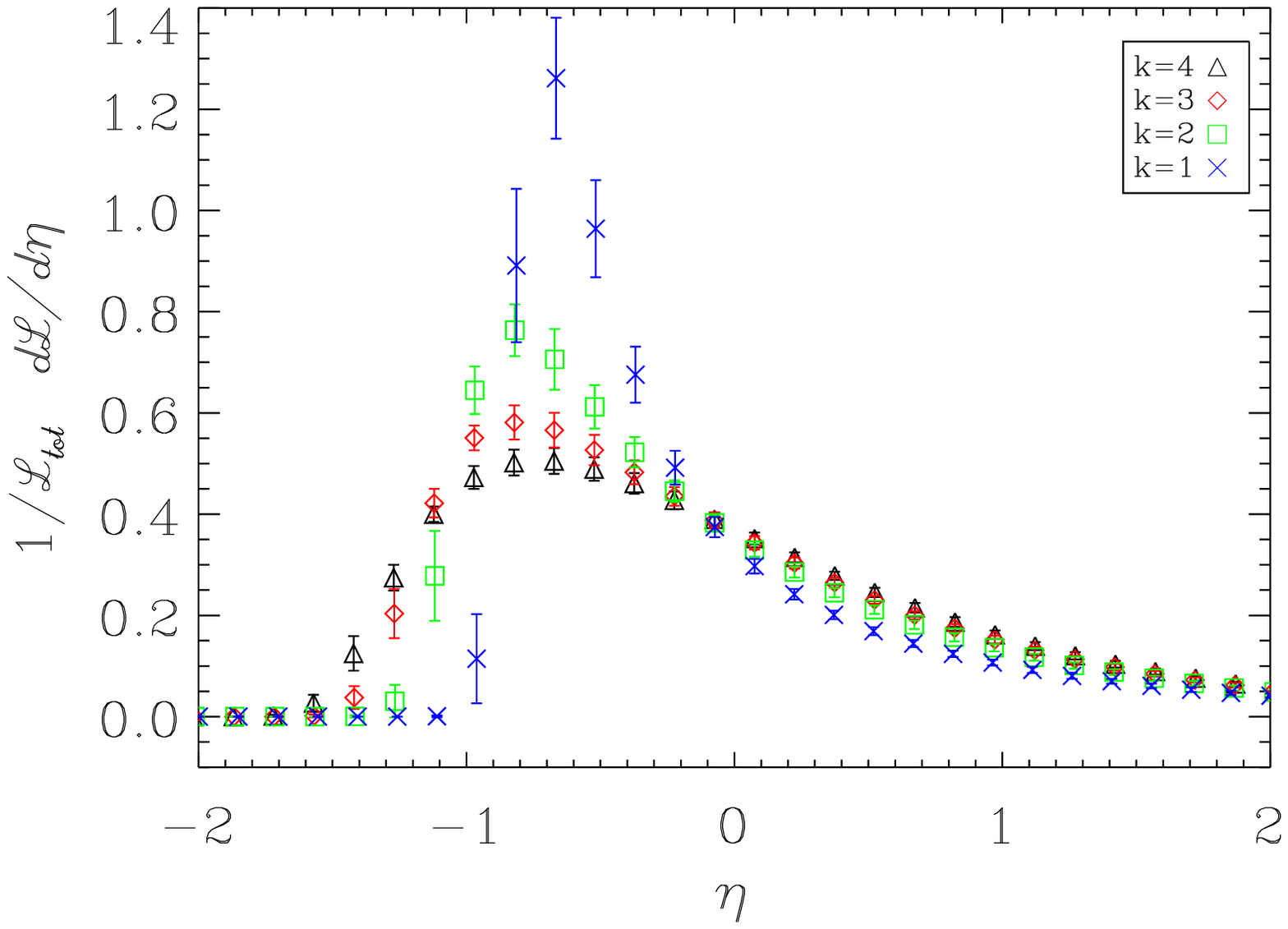,bbllx=66pt,bblly=364pt,bburx=539pt,bbury=700pt,width=8cm}
}\hbox{
\psfig{file=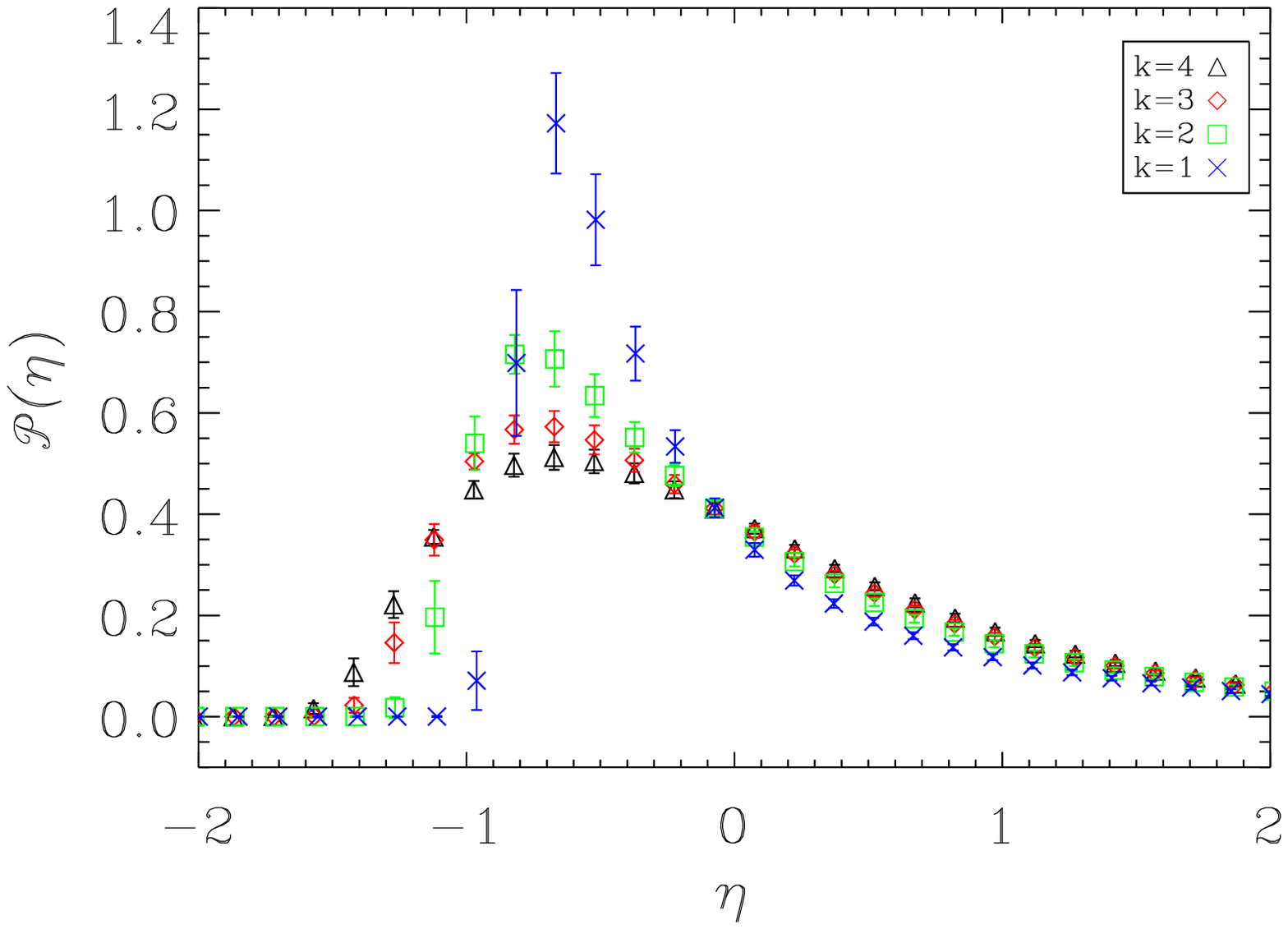,bbllx=66pt,bblly=364pt,bburx=539pt,bbury=700pt,width=8cm}
}} 
\caption[]{The measured differential length (left panel) of the total local skeleton 
[eq.~(\ref{eq:localsketot})] and the
pdf of the smoothed field (right panel) for
$\chi^2$ distributions with $k=1$, $2$, $3$ and $4$ degrees of freedom as
indicated on the upper right part of each panel.}
\label{figure9}
\end{figure*}
We now consider a few non Gaussian experiments. 
The first one is the case of a $\chi^2$ distribution with $k$
degrees of freedom (using scale-free Gaussian seeding fields with
spectral index $n=-2$). 
Fig.~\ref{figure9} is similar to Fig.~\ref{figure8}, but for $\chi^2$
distributions with $k=1$,  $2$, $3$ and $4$. The number of realizations, the
resolution of the maps and the smoothing are the same as in Fig.~\ref{figure8}. 
Clearly, the non Gaussian nature of the field is well reflected by the
skeleton. In fact, and quite surprisingly, its 
differential length again scales very much like the pdf of the smoothed
density field: there is very little difference between left and right
panels of Fig.~\ref{figure9}, except maybe for the size of the
errorbars: those are slightly larger for the skeleton than for the pdf,
as expected. Indeed, the skeleton construction relies on estimates of
derivatives in a one dimensional subset of pixels in the density map,
it is therefore more sensitive to noise. 

\begin{figure}
\centerline{\hbox{
\psfig{file=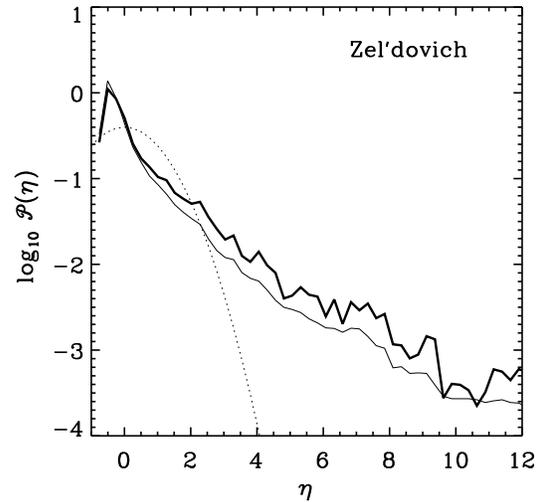,bbllx=52pt,bblly=117pt,bburx=426pt,bbury=485pt,width=7cm} 
}}
\caption[]{Comparison of differential length of the total local
skeleton [points of space verifying eq.~(\ref{eq:localsketot})] with
the pdf of the smoothed density field for the Zel'dovich map of
Sect.~\ref{sec:skedef2d}, both as functions of the normalized density contrast
$\eta=\delta/\sigma$. The thick/thin curve corresponds to the
skeleton/pdf. For reference, the Gaussian limit is also plotted as a dotted curve.}
\label{figure10}
\end{figure}

Our second non Gaussian experiment is the Zel'dovich map studied
in \S~\ref{sec:skedef2d} (Fig.~\ref{figure3}). Figure~\ref{figure10} compares
the differential length of the total local skeleton to the measured pdf. 
Again, for this single realization of a strongly non Gaussian field, 
the agreement between both measurements is very good, even in the
high density tail. Note that the curve for the skeleton is slightly
more irregular than the one for the pdf, as expected.

\begin{figure}
\centerline{\hbox{
\psfig{file=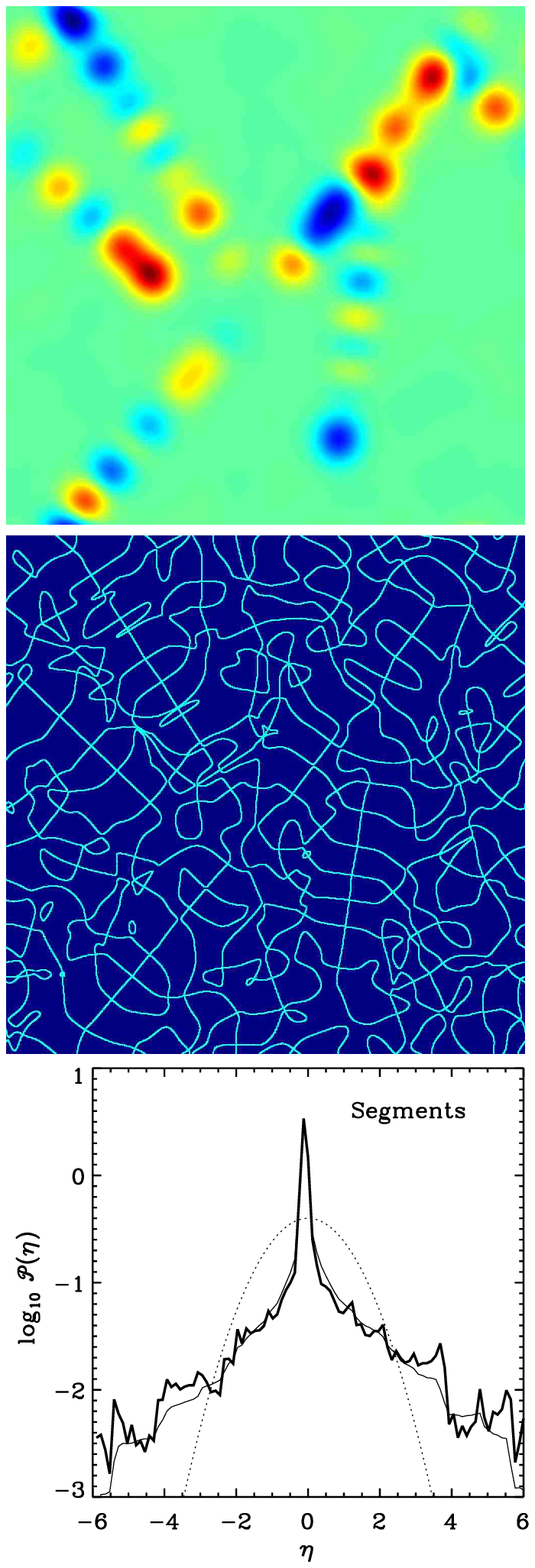,bbllx=211pt,bblly=167pt,bburx=421pt,bbury=763pt,width=6.5cm}
}}
\caption[]{The case of a Gaussian field with lines superposed on it.

{\em Top panel:} the Gaussian field of left panel of
Fig.~\ref{figure0}, where the density contrast has been increased by
a factor 1600 along five lines (800 pixels long and one pixel large)
randomly located and oriented, then smoothed
with a Gaussian window of radius 30 pixels. 

{\em Middle panel:} the total local skeleton obtained from the
smoothed field [points of space verifying eq.~(\ref{eq:localsketot})].

{\em Bottom panel:} the measured skeleton length (thick line) compared to the
pdf (thin line) of the smooth density field, as functions of the normalized
density contrast $\eta =\delta/\sigma$. The dots correspond to the
Gaussian limit. Note that due to our procedure, the field has very
strong kurtosis, but no skewness. Note as well that the central peak
is a Gaussian with the variance of the initial Gaussian field of
right panel of Fig.~\ref{figure0}.}
\label{figure11}
\end{figure}

Finally, to confirm the validity of the striking results of this
section, we decided to perform a quite extreme test as illustrated
by Fig.~\ref{figure11}. Taking the field generated on left panel of
Fig.~\ref{figure0}, we increased locally the density contrast by a
factor 1600 along 5 lines 800 pixels long and 1 pixel large, with
random positions and random orientations. This map was then smoothed with a
Gaussian window of radius 30 pixels, as shown on upper panel of Fig.~\ref{figure11}. The total
local skeleton obtained from this map is displayed in middle panel. The lines are
clearly visible on the picture as comb like structures. As shown in
lower panel, the skeleton differential length again scales very much like the pdf. 

\section{Discussion and links to dynamics}
\label{sec:discu}

In this paper, we studied some properties of the skeleton of a 2D random smooth field. This latter
is given by an ensemble of special field lines connecting saddle points to extrema. It 
is aimed to give accurate account of the network of filaments in the field.
The skeleton is a nonlocal object, difficult to build with a reliable algorithm
and to use to perform analytic calculations. We thus tried to find a local approximation to it, depending
on the field and its derivatives. To do that, we used two approaches, a mathematically motivated one
based on a Taylor expansion of the field around local maxima and saddle points, and a physically 
motivated one based on an examination of isocontour lines. They both lead to the same conclusion.
After comparing the resulting {\em local} skeleton to the real one, we performed statistical 
analyses of its length as a function of density threshold. To simplify the calculations, 
we considered a larger set of curves than the local skeleton but still containing it, that we called the 
{\em total} local skeleton. Having initially in mind to use the total local skeleton as
a test of non Gaussianity in CMB maps, we compared its differential length as function of the density 
threshold with the measured probability distribution function (pdf) of the smoothed field. 
The results of our paper can be summarized as follows:
\begin{enumerate}
\item By definition the real skeleton is the ensemble of pairs of field lines 
departing from saddle points, aligned initially with the major axis of local curvature
(corresponding to the largest eigenvalue of the Hessian)
and connecting them to local maxima. These field lines are drawn by going
along the trajectories with the equation of motion $d\vr/dt=\nabla \rho$ until convergence
to a local maximum. 
\item A very good approximation to the real skeleton, the {\em local} skeleton, 
is given by points of space
where the gradient is aligned with the major axis of local curvature and where the second component
of the local curvature is negative (i.e. the smallest eigenvalue of the Hessian is negative). 
We noticed however that the local skeleton was shorter than the real one, as expected, due to
the Lagrangian nature of the latter, which can have more than two fields lines converging to
local maxima, at variance with the former. 
\item The {\em total} skeleton is given by all the points of space where the gradient
is aligned with one axis of the curvature. Its differential length, as a function of
density threshold, is seen to scale very closely like the pdf of the smoothed density field.
This is explicitly demonstrated in the Gaussian case with analytic calculations. 
\end{enumerate}
The result mentioned in last point might be discouraging for using the total local skeleton as a test of non Gaussianity in 2D
maps, since it does not do better than the much simpler pdf.  Moreover, since the skeleton 
depends on local derivatives of the density field, it is expected to behave less well than the pdf 
with respect to the noise, although we did not investigate that in details, except in part for
cosmic variance effects. However, our analysis was not exhaustive. There might exist 
some counterexamples where the skeleton differential length scales differently from the pdf.
Our analyses relied on isotropic smoothing with a Gaussian window and we suspect that this contributes
to making the skeleton differential length very alike the pdf. 
More importantly in our analyses, we kept the full skeleton length, 
${\cal L}(-\infty)$, as an adjustable parameter. In fact, this 
length does bring significant additional pieces
of information and should be taken into account while comparing 
models predictions to measurements. 
Its analytical calculation in the Gaussian limit, although theoretically feasible, is
rather cumbersome: we left it for future work, having in mind that it can be fairly determined
numerically through appropriate realizations of the models.

Along this paper, we did not address dynamics, although we used a Zel'dovich map
to illustrate our purpose. The skeleton is in fact a quite useful tool for the
analysis of the large scale structure distribution and the understanding of its dynamics.
Indeed, recall that we initially defined the skeleton as
the border of void patches (\S~\ref{sec:definition}) and 
that the void patches are given by the regions of space containing all the points converging 
to the same local minimum while going along the field lines
in opposite direction to the gradient. On can similarly associate
the peak patches to local maxima (Bond \& Myers 1996a). 
Together with the skeleton, 
the peak patches are the building blocks of the
observed large scale structures in the Universe. 

\begin{figure*}
\centerline{\hbox{
\psfig{file=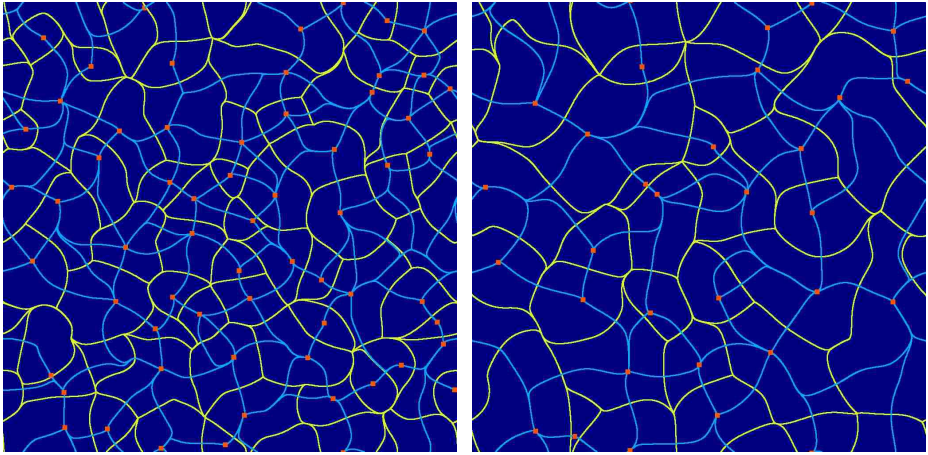,bbllx=116pt,bblly=562pt,bburx=523pt,bbury=763pt,width=12.5cm}
}}
\caption[]{The skeleton and the peak patches for the Gaussian field of
Fig.~\ref{figure0}, at two different smoothing radii $\ell$. 

The left panel corresponds to $\ell_1=25$ pixels, as previously. The right panel
corresponds to $\ell_2=25 \sqrt{2}\simeq 36$ pixels. The skeleton is
represented by the blue lines, and the peak patches borders (the dual
skeleton) by the golden ones. Each red point corresponds to a local
maximum.  Left and right panels can be seen
as a Lagrangian view of the system at two different times, describing
the merging of structures. 

Clearly, the peak patches of
the right panel can be seen as mergers of peak patches of the left
panel, even if mergers can occur differently according to the place of
interest: some peak patches survive, i.e. do not merge with others,
some of them experience merging with one or more neighbors. 
As a result, the skeleton of right panel is approximately
made of a connected subset of lines composing the skeleton  of left
panel.

Note that on the lower left corner of left panel, 
there is a peak patch containing no local
maximum. This is clearly an artifact from our numerical approach,
which did not detect it. This is not surprising since we 
noticed earlier (Fig.~\ref{figure1}) 
that something was wrong with the connectivity of the skeleton between
critical points in this
location.}
\label{figure6}
\end{figure*}

Indeed, in the standard approach of hierarchical formation of
galaxies, the peak patches collapse and merge together at successive
times. The merging of peak patches can also be seen as the
collapse of larger peak patches constituted by their progenitors. 
Within a multi-scale approach, these latter can be obtained by
smoothing the field at increasing scales, each smoothing scale
corresponding to a different collapse time, as illustrated
by Fig.~\ref{figure6}. This peak patch approach was in fact
used extensively by Bond \& Myers (1996a,b,c)  to produce a
simplified but quite accurate description of large structure 
dynamics, including the evolution of non linear
objects from galaxies to clusters of galaxies, their merging
history as well as their large scale motions.\footnote{See also Monaco
et al. 2002 for a similar algorithm.} 
Note that peak patches are rather compact, and
thus can be well approximated by ellipsoids, lending credence to the
Bond \& Myers approach. This also stems from a Taylor
expansion of the field around local maxima. 

This line of thought can be followed further. Indeed, the local maxima
are by definition located on the skeleton along which the matter
flows: merging of earlier collapsed patches will take place
at the nodes of the skeleton (see also caption of
Fig.~\ref{figure6}). This description is well known and
understood through the adhesion approximation (e.g., Kofman \& Shandarin 1988;
Kofman, Pogosyan \& Shandarin 1990). 

\begin{figure*}
\centerline{\hbox{
\psfig{file=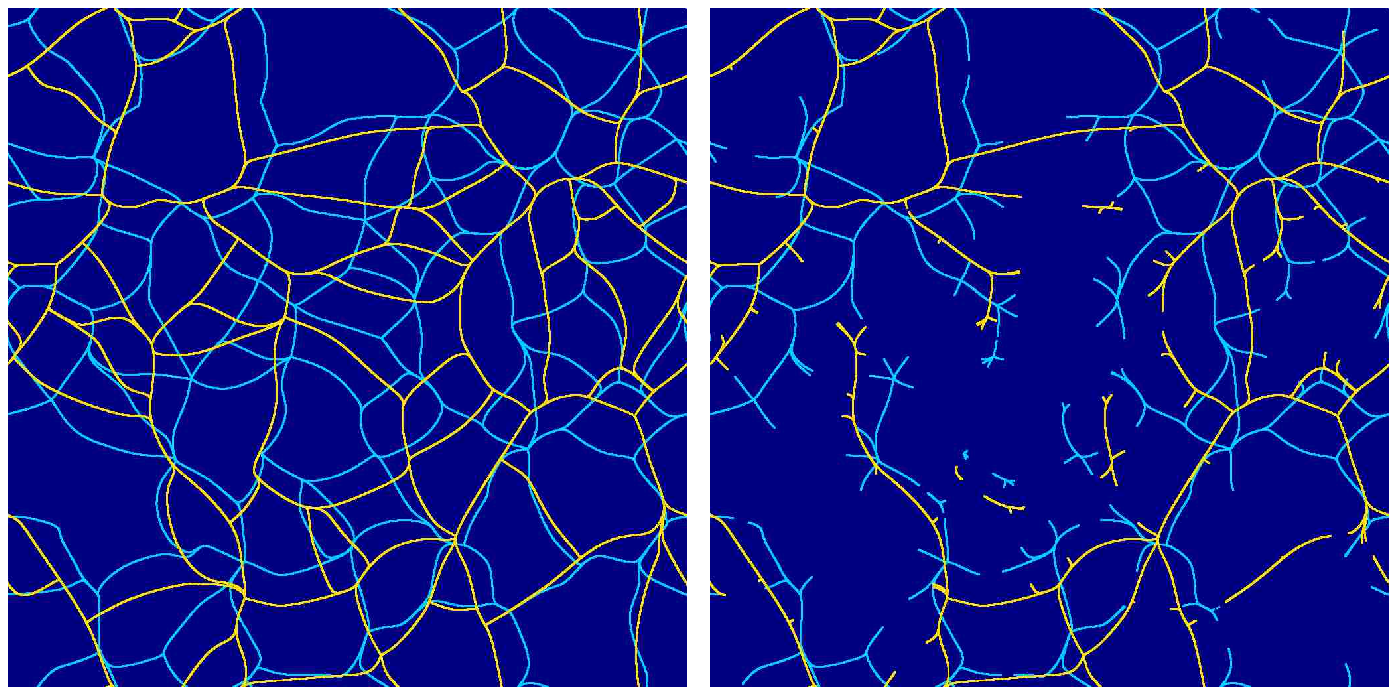,bbllx=116pt,bblly=562pt,bburx=523pt,bbury=763pt,width=12.5cm}
}} 
\caption[]{Dynamical evolution of the skeleton for the smooth Gaussian
field of Fig.~\ref{figure0}.

{\em Left panel:} the final skeleton obtained from the Zel'dovich mover (i.e. as
explained in Fig.~\ref{figure3}, golden lines) superposed to the
initial one (blue lines).

{\em Right panel:} same as left panel but for over-dense regions. 

One can clearly match the details in the initial pattern of filaments to the final one,
except for mergers as in e.g. the top right corner and the left side
of the panel. Note as well the expansion of the central under-dense region.}
\label{figure7}
\end{figure*}

Note however that the skeleton itself has its own dynamics, as
illustrated by Fig.~\ref{figure7}: filaments composing it can move and
be distorted due to large scale flows, but can also merge together. For
the particular example considered here, Fig.~\ref{figure4} shows that
the total length of the skeleton is approximately conserved. It is
slightly shorter in the Zel'dovich case compared to initial conditions
as a result of the competition between merging and stretching.  

Since the matter tends to flow along the lines of the skeleton, these
latter represent a useful reference frame to study internal structure
and dynamics of filaments in the Universe, with well prescribed 
procedure to define them, given a typical scale length. In practice, 
this latter should be larger than the size of clusters of galaxies, and smaller
than the size of super-clusters, i.e. of order of a few Mpc. In that
case, one knows that the skeleton reflects rather well the initial
primordial one (Bond, Kofman \& Pogosyan 1996). Note again that the skeleton
length can be measured in a cosmological volume, and compared to
theoretical predictions. 

Since we are in the 2D case, the discussion concerning the dynamics remained at 
the qualitative level. More quantitative analyses in 3D are left for future work. 
 
\section*{Acknowledgments}

S.C. and O.D. thanks the Oxford Astrophysics group, where part of this
project was conducted, for its hospitality. We also thank J.~R. Bond, F.~R. Bouchet and
D. Pogosyan for useful discussions. The numerical calculations were performed with a dedicated
\texttt{FORTRAN 90} package, {\bf \texttt{FasToCh}}, available on request from the authors and
developed as part of GPH428 submodule of HFI level 2 in the framework of Planck Surveyor preparation. 
The numerical calculations were performed
at IAP on MAGIQUE (Silicon Graphics O3K, 14 processors, 20Gb memory).



\appendix

\section{Numerical approach}

The numerical calculations were performed with a dedicated
\texttt{FORTRAN 90} package: {\bf \texttt{FasToCh}}\footnote{Available
on request from the authors.} ({\bf Fas}t {\bf
To}pological {\bf Ch}ase). 

While confronting the measurements to the theoretical predictions of Sect.~\ref{sec:gaussian} 
for the Gaussian case, we tested extensively pixelization and finite volume effects
by generating maps at various resolutions and smoothed at various scales. We
tried as well various schemes described below for computing derivatives 
and interpolating the field, and we run many different realizations of the same 
power-spectrum. Finite volume effects 
are more important for smaller values of the spectral index, 
$n$, while pixelization effects, on the contrary, increase with $n$. In principle if
the smoothing length 
$\ell$ is very large compared to the pixel size and very small compared to the map size, $L$,
both these effects should be negligible, as we found
for  $n=0$, $-1$ and $-2$. Practically, 
$10$ pixels $\la \ell \la L/20$ is generally a
safe choice, but in fact the range of available scales depends on 
the statistics considered as discussed more in detail below an on 
details of the field properties, in particular its power-spectrum shape. 
We also tested anisotropy effects by
generating the initial random field with or without Hanning filtering, 
the latter insuring isotropy at  small scales (e.g. Berstchinger 2001) and found
that they were insignificant.

Before entering into the details of the skeleton construction, we first
detail the issue of computing reliably the successive derivatives of
the field. We examined both Fourier methods and the simplest finite
difference schemes. Throughout this paper, we used the Fourier method
but the simple finite difference method is both faster and easier to
use when the coverage is more intricate such as in galaxy catalogs or
in CMB experiments. We also investigated pixelization effects and found that 
they are negligible provided that the smoothing window is large enough, i.e.
a few pixels radius, typically $\ell \ga 3$. 
However this of course depends on the type
of unsmoothed map considered: if there is a lot of small scale power 
in the map, it is necessary to smooth it more to have reliable estimates 
of the derivatives. 

We also tested bicubic interpolation (e.g. Press et al. 1992) which
has the advantage of warranting  divergence-free gradient at all
positions within a pixel. Practically, this means
that given the field values at the pixel centers, as well as
pre-computations of the gradient and of the diagonal terms of the
Hessian (with either Fourier or finite difference), any quantity can be computed
self-consistently at any location within the pixels, including the off-diagonal terms
of the Hessian. This property is in principle particularly critical when building the
real skeleton. In practice, however, the improvements brought by the
bicubic interpolation were insignificant as compared to the simpler and
faster bilinear interpolation, which was finally used for all the calculations.

Drawing the local skeleton is simple if one sees that the
equation ${\cal S} = 0$ [eq.~(\ref{eq:localsketot})]
corresponds to the zero isocontour of the field  ${\cal S}$. This can be performed
with a standard method as we now explain. 
Given a square of four neighboring pixels and the corresponding values of ${\cal
S}$, we first determine whether the linearly interpolated field
cancels along two sides of the square. If this
happens, this means that the isocontour curve crosses the square. We
locally approximate this curve by a segment which extremities are
located on the edges of the square. The coordinates of the segment
ends are easily found using dual interpolation. 
The length of the skeleton is found by adding all the individual 
segment lengths. There are however particular cases where two pieces of
isocontour can intersect at the same point (e.g., at a critical
point) or become very close to each other. 
This can produce configurations where the field cancels on
4 edges of the square. In that case, we cannot 
compute reliably the length of the
isocontour within the square, but the relative contribution of these 
configurations is increasingly small with the smoothing length. To make these
contributions negligible, and to insure as well that 
the effect of approximating the local skeleton locally by straight lines	
is negligible,  the smoothing length should be of order a few pixels size,  
typically $\ell \ga 5$ for spectral index $n \la -1$. 
However, the smoothing radius has to remain small compared to the 
map size, in order to avoid finite volume effects, typically $\ell \la L/20$, where
$L$ is the map size.

Drawing the real skeleton is rather difficult, at least we did not
find yet a highly reliable algorithm. 
As explained in Sect.~\ref{sec:definition} the real skeleton is drawn
by going along the trajectory with the following motion equation
\begin{equation}
\frac{d \vr}{dt}\equiv \vv=\nabla \rho, \label{eq:motion2}
\end{equation}
starting from the saddle points and with initial velocity parallel to
the major axis of the local curvature (the dual skeleton is obtained
similarly by making the operation $\rho \rightarrow -\rho$). 
This equation of motion is solved numerically with a semi-implicit
scheme which guaranties that the gradient does not change sign along
the direction of motion. This allows as a consequence a correct
calculation of the skeleton length since backwards motion
along the trajectory is prevented.

More explicitly, if $\vr_i$, $\vv_i=\nabla \rho_i$ and
$dt_i$ are respectively the position, velocity and timestep at step
$i$, the quantities at next step are computed as follows using an
iterative procedure. The gradient $\nabla \rho$ is evaluated at the position ${\tilde
\vr}_{i+1}=\vr_i+{\tilde \vv}_{i+1} d{\tilde t}_{i+1}$ from which we
deduce ${\tilde \vv}_{i+1}$ and $s={\tilde \vv}_{i+1}.\vv_i$. As
a first guess we take $d{\tilde t}_{i+1}=dt_i$ and ${\tilde \vv}_{i+1}=\vv_i$.
As long as $s\leq 0$ we divide the timestep 
$d{\tilde t}_{i+1}$ by two and recompute ${\tilde \vr}_{i+1}$.
Once $s>0$, the quantities at time step $i+1$ are set to
$dt_{i+1}=d{\tilde t}_{i+1}$, $\vv_{i+1}={\tilde \vv}_{i+1}$ and
$\vr_{i+1}=\vr_i+{\tilde \vv}_{i+1} {\tilde dt}_{i+1}$.  Note in addition that
the time step is always chosen such that the displacement between two steps is at 
most of the order of a pixel.

To ensure a finite number of
time steps, the motion is stopped once we reached the vicinity
(typically a fraction of pixel size) 
of a critical point and appear to converge
towards it. In principle this stopping point should always be a
maximum. However, in practice we found that very rarely we could
converge to a saddle point. To avoid here again an infinite number of iterations,
we stop the calculation of the trajectory. Indeed, this situation happens when
a field line (say $A$) is very close to one of the 
two unstable field lines converging to the saddle point
(see Fig.~\ref{figure2}). After getting close to the saddle point, the field line $A$
should turn by approximately 90 degrees to follow closely one of the two stable field lines
(say $B$) of the saddle point. The field line $B$ will be drawn anyway starting from this 
saddle point. So visually, we should not miss anything by not drawing the end of the field
line $A$. However, this might be a problem for
estimating the skeleton length since we are missing parts of it. 

Eventually, the critical points are determined as intersections of the
contour lines $\partial \rho/\partial r_1 = 0$ and  $\partial
\rho/\partial r_2 = 0$ with the same method as used for the local
skeleton. This detection method can also fail (e.g. upper left panel of 
Fig.~\ref{figure1} and \ref{figure6}), 
and some critical points
can be missing, which has dramatic consequences for the building of the
real skeleton.

To avoid critical situations as described above, where our algorithm for drawing
the real skeleton fails, we need to smooth
the fields significantly, typically with a radius at least of order of 10 pixels
but this depends strongly on the nature of the field.


\begin{thebibliography}{}
\bibitem{}  Aghanim, N., Forni, O., 1999, A\&A 347, 409
\bibitem{}  Bernardeau, F., Colombi, S., Gazta\~naga, E., Scoccimarro, R., 2002, Phys. Rep. 367, 1
\bibitem{}  Babul, A., Starkman, G. D., 1992, ApJ 401, 28
\bibitem{}  Bardeen, J. M., Bond, J. R., Kaiser, N., Szalay, A. S., 1986, ApJ 304, 15
\bibitem{}  Barrow, J. D., Bhavsar, S. P., Sonoda, D. H., MNRAS 216, 17
\bibitem{}  Bertschinger, E., 2001, ApJS 137, 1
\bibitem{}  Bhavsar, S. P., Barrow, J. D., 1983, MNRAS 205, 61P
\bibitem{}  Bond, J. R., Efstathiou, G., 1987, MNRAS 226, 655
\bibitem{}  Bond, J. R., Kofman, L., Pogosyan, D., 1996, Nature 380, 603
\bibitem{}  Bond, J. R., Myers, S. T., 1996a, ApJS 103, 1 
\bibitem{}  Bond, J. R., Myers, S. T., 1996b, ApJS 103, 41 
\bibitem{}  Bond, J. R., Myers, S. T., 1996c, ApJS 103, 63 
\bibitem{}  Chiang, L.-Y., Coles, P., 2000, MNRAS 311, 809
\bibitem{}  Colombi, S., Pogosyan, D., Souradeep, T., 2000, Phys. Rev. Lett. 85, 5515
\bibitem{}  Dor\'e, O., Colombi, S., Bouchet, F. R., 2003, MNRAS, in press 
            (astro-ph/0202135)
\bibitem{}  Doroshkevich, A. G., 1970,  Astrofizica, 6, 581 [Astrophysics 6, 320]
\bibitem{}  Doroshkevich, A. G., Tucker, D. L., Lin, H., Turchaninov, V., Fong, R., 2001, MNRAS 322, 369
\bibitem{}  Gott, J. R. III, Melott, A. L., Dickinson, M., 1986, ApJ 306, 341
\bibitem{}  Hobson, M. P., Jones, A. W., Lasenby, A. N., 1999, MNRAS 309, 125
\bibitem{}  Jost, J., 2002, Riemannian Geometry and Geometric Analysis (Springer, third
            edition)
\bibitem{}  Kerscher, M., 2000, Lecture Notes in Physics 554,  36
\bibitem{}  Kofman, L., Shandarin, S., 1988, Nature 334, 129
\bibitem{}  Kofman, L., Pogosyan, D., Shandarin, S., 1990, MNRAS 242, 200
\bibitem{}  Mecke, K. R., Buchert, T., Wagner, H., 1994, A\&A 288, 697
\bibitem{}  Milnor, J., 1963, Morse Theory (Princeton University, Princeton, NJ)
\bibitem{}  Monaco, P., Theuns, T., Taffoni, G., Governato, F., Quinn, T., Stadel, J., 2002, ApJ 564, 8
\bibitem{}  Naselsky, P. D., Novikov, D. I., Silk, J., 2002, ApJ 565, 655
\bibitem{}  Peebles, P. J. E., 1980,  The Large-Scale Structure of the Universe (Princeton Univ. Press, 1980)
\bibitem{}  Phillips, N. G., Kogut, A., 2001, ApJ 548, 540
\bibitem{}  Press, W. H., Teukolsky, S. A., Vetterling, W., Flannery, B. P., 1992, 
            Numerical Recipes in FORTRAN. The Art of Scientific Computing 
            (Cambridge Univ. Press, 1992)
\bibitem{}  Sahni, V., Sathyaprakash, B. S., Shandarin, S. F., 1998, ApJ 495, L1
\bibitem{}  Shandarin, S. F., 1983, Sov. Astron. Lett. 9, 104
\bibitem{}  Shandarin, S. F., 2002, MNRAS 331, 865
\bibitem{}  Stuart, A., Ord, J. K., 1994, Kendall's Advanced Theory of Statistics, sixth edition,
            volume I, Distribution Theory (Edward Arnold, Halsted Press, New York-Toronto)
\bibitem{}  Tomita, H., 1986, Prog. Theor. Phys. 75, 482
\bibitem{}  Zel'dovich, Ya. B., 1970, A\&A 5, 84
\bibitem{}  Zel'dovich, Ya. B., 1982, Sov. Astron. Lett. 8, 102
\bibitem{}  Zel'dovich, Ya. B., Einasto, J., Shandarin, S. F., 1982, Nature 300, 407




\end{thebibliography}
\end{document}